\documentclass[twocolumn]{aastex63}
\usepackage[normalem]{ulem}   
\usepackage{nameref}
\usepackage{hyperref}
\usepackage{amsmath}
\usepackage{amssymb}
\usepackage{mathrsfs}
\usepackage{animate}
\usepackage{graphicx}
\usepackage{rotating}
\usepackage{bbm} 
\usepackage{amsthm}        
\usepackage{comment}

\usepackage{svg}
\usepackage[outline]{contour}

\usepackage{subfigure}
\usepackage{tikz}
\usepackage{mdframed}

\usepackage{lineno}
\usepackage{multirow}

\newcommand{\Tp}{T_\mathrm{p}}
\newcommand{\Te}{T_\mathrm{e}}

\newcommand{\Brpsp}{B_\mathrm{r|PSP}}
\newcommand{\Brsolo}{B_\mathrm{r|SO}}
\newcommand{\Bropsp}{B_\mathrm{r0|PSP}}
\newcommand{\Brosolo}{B_\mathrm{r0|SO}}
\newcommand{\fpsp}{f_\mathrm{PSP}}
\newcommand{\fsolo}{f_\mathrm{SO}}
\newcommand{\thetapsp}{\theta_\mathrm{PSP}}
\newcommand{\thetasolo}{\theta_\mathrm{SO}}
\newcommand{\dtp}{\delta_{\theta, \phi}}
\newcommand{\Mf}{M_\mathrm{flux}}
\newcommand{\Mfpsp}{M_\mathrm{flux|PSP}}
\newcommand{\Mfsolo}{M_\mathrm{flux|SO}}
\newcommand{\tpsp}{t_\mathrm{PSP}}
\newcommand{\tsolo}{t_\mathrm{SO}}

\newcommand{\rpsp}{r_\mathrm{PSP}}
\newcommand{\rsolo}{r_\mathrm{SO}}

\newcommand{\finj}{f_\mathrm{inj}}
\newcommand{\Tinj}{\tau_\mathrm{inj}}
\newcommand{\rs}{r_\odot}
\newcommand{\fss}{f_\mathrm{ss}}
\newcommand{\rss}{r_\mathrm{ss}}

\newcommand{\gamp}{\gamma_\mathrm{p}}
\newcommand{\game}{\gamma_\mathrm{e}}

\newcommand{\Dv}{\Delta v}

\newcommand{\DTp}{\Delta \Tp}
\newcommand{\DTe}{\Delta \Te}
\newcommand{\Ddv}{\Delta \dv}
\newcommand{\Ddb}{\Delta \db}
\newcommand{\Dsigmac}{\Delta \sigma_c}
\newcommand{\Dnp}{\Delta n_p}

\newcommand{\dv}{\delta v}

\newcommand{\db}{\delta b}
\newcommand{\psp}{\mathrm{PSP}}
\newcommand{\so}{\mathrm{SO}}
\newcommand{\spani}{\mathrm{SPAN-I}}
\newcommand{\spc}{\mathrm{SPC}}

\newcommand{\rc}{r_\mathrm{c}}
\newcommand{\Css}{C_\mathrm{ss}}

\newcommand{\dvvect}{\delta \mathrm{\textbf{v}}}
\newcommand{\dbvect}{\delta \mathrm{\textbf{b}}}

\newcommand{\rev}{}

\shorttitle{Source Alignment Method}
\shortauthors{Dakeyo et al.}

\begin{document}

\title{On the Radial Evolution of the Solar Wind : The Source Alignment Method Applied to Parker Solar Probe and Solar Orbiter Observations}

\correspondingauthor{Jean-Baptiste Dakeyo}
\email{jbdakeyo@berkeley.edu}

\author[0000-0002-1628-0276]{Jean-Baptiste Dakeyo}
\affiliation{Space Sciences Laboratory, University of California, Berkeley, CA, USA}

\author[0000-0002-8475-8606]{Tamar Ervin}
\affiliation{Space Sciences Laboratory, University of California, Berkeley, CA, USA}
\affiliation{Department of Physics, University of California, Berkeley, CA, USA}

\author[0000-0002-1989-3596]{Stuart Bale}
\affiliation{Space Sciences Laboratory, University of California, Berkeley, CA, USA}
\affiliation{Department of Physics, University of California, Berkeley, CA, USA}

\author[0000-0001-8215-6532]{Pascal D\'emoulin}
\affiliation{LIRA, Observatoire de Paris, Universit\'e PSL, CNRS, Sorbonne Universit\'e, Universit\'e de Paris, 5 place Jules Janssen, 92195 Meudon, France}
\affiliation{Laboratoire Cogitamus, 75005 Paris, France}

\author[0000-0002-1128-9685]{Nikos Sioulas}
\affiliation{Space Sciences Laboratory, University of California, Berkeley, CA, USA}

\author[0000-0002-2916-3837]{Victor Réville}
\affiliation{IRAP, Universit\'e de Toulouse, CNRS, CNES, 9 Avenue du Colonel Roche, 31400 Toulouse, France}

\author[0000-0003-2981-0544]{Mingzhe Liu}
\affiliation{Space Sciences Laboratory, University of California, Berkeley, CA, USA}

\author[0000-0003-4039-5767]{Alexis Rouillard}
\affiliation{IRAP, Universit\'e de Toulouse, CNRS, CNES, 9 Avenue du Colonel Roche, 31400 Toulouse, France}
\affiliation{Leibniz-Institut für Astrophysik Potsdam, Potsdam, 14482, Germany}

\author[0000-0001-6172-5062]{Milan Maksimovic}
\affiliation{LIRA, Observatoire de Paris, Universit\'e PSL, CNRS, Sorbonne Universit\'e, Universit\'e de Paris, 5 place Jules Janssen, 92195 Meudon, France}

\author[0000-0001-5030-6030]{Davin Larson}
\affiliation{Space Sciences Laboratory, University of California, Berkeley, CA, USA}

\author[0000-0002-4559-2199]{Orlando Romeo}
\affiliation{Space Sciences Laboratory, University of California, Berkeley, CA, USA}

\author{Philippe Louarn}
\affiliation{IRAP, Observatoire Midi-Pyrénées, Universit\'e Toulouse III - Paul Sabatier, CNRS, 9 Avenue du Colonel Roche, 31400 Toulouse, France}

\author[0000-0002-0396-0547]{Roberto Livi}
\affiliation{Space Sciences Laboratory, University of California, Berkeley, CA, USA}

\begin{abstract}
The properties of the solar wind, as measured in situ throughout the heliosphere, depend both on the characteristics of its coronal source and on the intrinsic processes governing its interplanetary evolution. Recently, radial and Parker spiral alignment techniques have been applied to Parker Solar Probe (PSP) and Solar Orbiter (SO) observations to investigate the radial evolution of the same solar wind parcel. These studies have shown that the solar wind can undergo significant acceleration even beyond its primary acceleration region (i.e., above 15 solar radii). However, such radial and Parker spiral alignments are rare in practice, which limits the statistical significance and general applicability of the results.
We introduce a new source alignment technique designed to overcome these limitations. Using magnetic backmapping, we associate similar solar wind streams observed by the two spacecraft based on the proximity of their photospheric footpoints, combined with additional in-situ stream similarity criteria. Applying the source alignment method to PSP and SO observations, we identify a total of 548 alignment intervals, each lasting 30 minutes. By constructing statistics over all alignments, we find that the solar wind speed increases by an average of 45\% \rev{per radial decade} (approximately 147 km/s) between the two probes. This result demonstrates that solar wind acceleration in the inner heliosphere remains significant compared to that occurring below 15 solar radii. Among the different \rev{studied plasma parameters}, the radial evolution of the electron \rev{temperature and plasma density,} show the strongest \rev{anti-}correlation with the increase in \rev{bulk velocity}. 
\end{abstract}


\section{Introduction}

Statistical analyses have been widely used to study the radial evolution of the solar wind \citep{maksimovic2020, dakeyo2022, Halekas2023}. One existing approach consists of defining wind speed populations. This allows the radial evolution of slow and fast solar wind streams to be studied separately, together with their respective plasma properties, such as the proton bulk speed $v_p$, the proton temperature $\Tp$, the electron temperature $\Te$, and the plasma density $n$. Numerous observations at 1~au have shown that $v_p$ is correlated with $\Tp$, partially correlated with $\Te$, and anticorrelated with $n$ at Earth’s orbital distance \citep{Lopez1986solar, Demoulin2009T-Vcorrelation, Elliott2012temporal, maksimovic2020}. However, these correlations do not remain fixed as the solar wind expands. Indeed, observations from \textit{Helios} \citep{1981_ref_helios} and Parker Solar Probe \citep[PSP;][]{fox2016solar} have shown that $v_p$ becomes anticorrelated with $\Te$ when observed closer to the Sun \citep{maksimovic2020} ($\lesssim$ 0.5~au). Therefore, plasma parameters and their correlations are not necessarily preserved during radial expansion.

Moreover, while statistical studies incorporate large amounts of data, they also rely on the implicit assumption that wind speed populations (from slow to fast) do not globally mix during their radial evolution. Although $v_p$, $\Tp$, $\Te$, and $n$ are strongly related \citep{maksimovic2020}, correlations alone do not guarantee that different wind populations remain unmixed, which may affect the physical interpretation of the results.

It is often assumed that knowledge of the properties of a solar wind stream observed \textit{in-situ} can help to infer plasma conditions in the inner heliosphere. However, a recent magnetic connectivity study applied to Solar Orbiter \citep[SO;][]{muller2020} observations, which related \textit{in situ} measurements to coronal wind characteristics (e.g., expansion factor and photospheric magnetic field intensity), showed that coronal properties alone are not sufficient to predict the radial evolution of the solar wind in most cases \citep{dakeyo2024b}. The complexity of solar wind radial evolution therefore makes it difficult to identify the key parameters governing this evolution.

Two recent inner heliospheric missions, PSP and SO, orbiting the Sun at different radial distances, have opened the possibility of following the radial evolution of a single solar wind plasma parcel during so-called “radial alignment” and “Parker spiral alignment” conjunctions. 
These techniques aim to track the same parcel of plasma using two spacecraft. Both are based on predicting the trajectory and time of travel, either using the purely radial expansion of the plasma (i.e. radial alignment) or the trajectory described by the Parker spiral in the Sun's rotating frame (Parker spiral alignment). 
Several studies have analyzed the radial evolution of solar wind properties during these conjunctions \citep{Telloni2021, Berriot2024, Berriot2025, Ervin2024a, Ervin2024b, Rivera2024, Rivera2025}, with the most recent results suggesting that Alfvén waves may heat and accelerate exclusively fast solar wind streams. These techniques allow the radial evolution of the solar wind to be studied while minimizing the mixing of different stream properties, thereby providing valuable insight into the underlying \textit{in situ} physical processes. However, due to the small number of available alignments and the difficulty of confidently identifying the same plasma parcel at the two probes, it is not yet possible to obtain global results regarding solar wind acceleration and the associated energy budget.

To overcome the limited occurrence of radial alignments, we develop a “source alignment” technique that we apply between PSP and SO. Source alignment is conceptually similar to Parker spiral alignment but focuses on identifying conjunctions of similar plasma parcels originating from the same source, rather than tracking the same plasma parcel between the two probes. This conceptual change in the stream association method also enables the radial evolution of the solar wind to be related to its source properties. In the present study, we investigate solar source alignments between PSP and SO to statistically study solar wind acceleration and the radial evolution of plasma properties.

In Section~\ref{sec:method}, we explain the concept of the source alignment method and its specific application to solar wind sources and \textit{in situ} properties. Section~\ref{sec:data_preprocess} presents the dataset used for this study and describes the preprocessing steps. Section~\ref{sec:stream_assoc_result} shows the results of the identified alignments between PSP and SO and the corresponding evolution of the main \textit{in situ} properties. Section~\ref{sec:correl_results} presents correlations between wind properties at PSP and SO during radial evolution. Finally, Section~\ref{sec:conclusion} summarizes the results and discusses their broader implications.

\section{Source Alignment Method}\label{sec:method}

\begin{figure*}[t!]
    \centering
    \includegraphics[width=16cm]{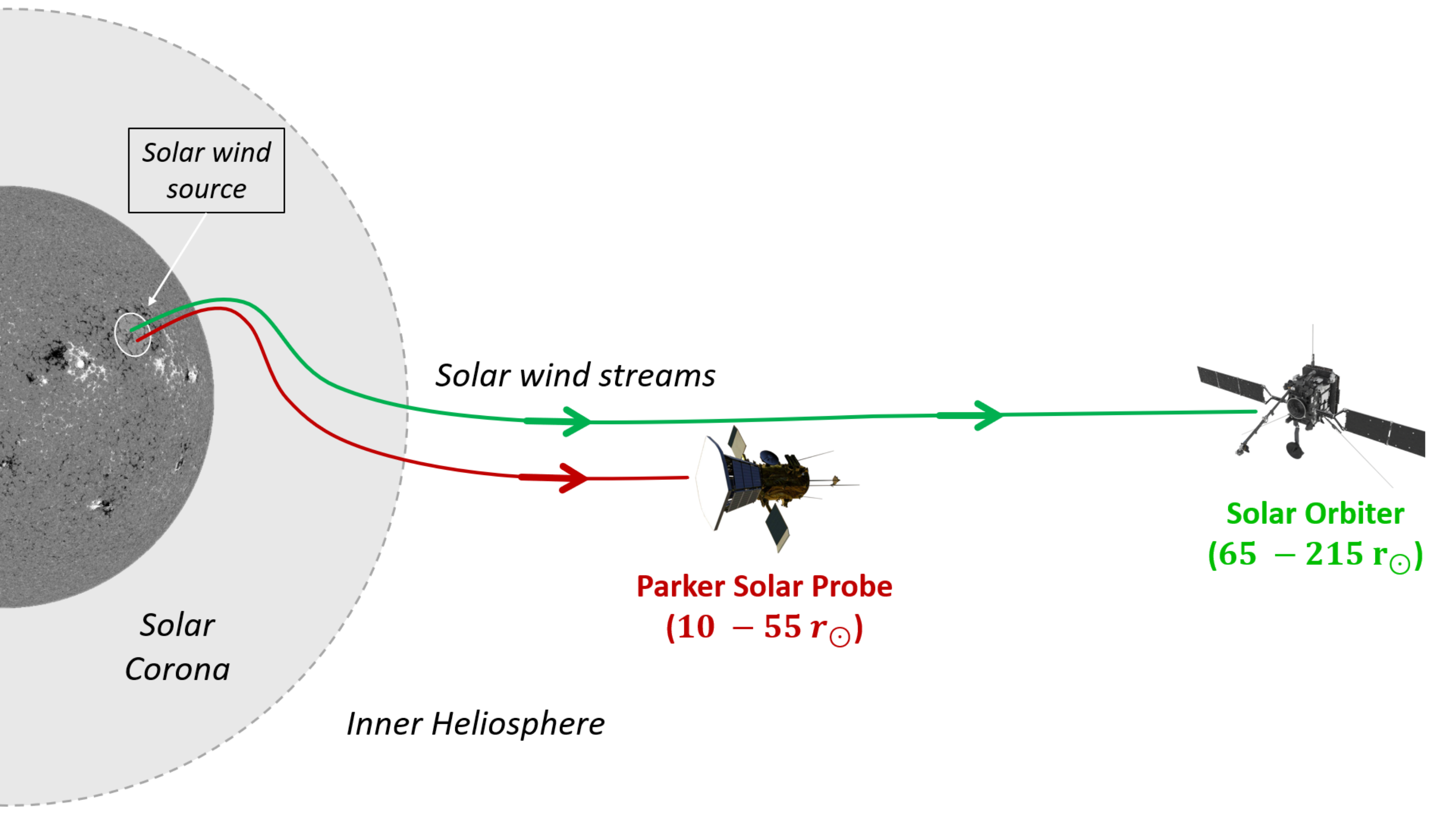}
    \caption{Schematic of a source alignment conjunction between PSP and SO. The map displayed as example on the illustration of the Sun, is one of the ADAPT magnetogram used in the study \citep{adapt_ref2013}. }
    \label{fig_illustr_source_align}
\end{figure*}

The source alignment method aims to identify similar solar wind streams observed by different spacecraft based on the locations of their respective magnetic footpoints in the solar photosphere (illustrated in Figure~\ref{fig_illustr_source_align}). Radial and Parker spiral alignment methods also seek to identify a given solar wind stream at multiple spacecraft locations. However, it has become clear that such alignments are difficult to detect in practice when building statistics on the radial evolution of individual streams. In particular, the spacecraft must follow closely similar heliolatitudes, and the alignment intervals typically last only tens of minutes to about an hour \citep{Berriot2024, Ervin2024b, Rivera2024, Rivera2025}. In addition, as the radial separation between the spacecraft increases, the plasma parcel undergoes stronger expansion and evolution, making it increasingly difficult to conclude that the same stream has been identified by both probes \citep{Berriot2024, Berriot2025}.

Unlike radial or Parker spiral alignments, the source alignment method does not assume that the same plasma parcel is observed by both spacecraft. Instead, it assumes that distinct plasma parcels originating from the same coronal source share similar intrinsic properties. These parcels may in some cases correspond to the same stream, but this is not a requirement. Moreover, a given solar wind source can remain stable for days or even weeks, while a spacecraft may remain magnetically connected to that source for hours or days. Since the properties of a solar wind source are expected to be nearly homogeneous, it is expected to drive wind streams of similar nature, then a single source is supposed to generate comparable solar wind streams throughout the inner heliosphere.

Source association therefore provides an effective way to substantially increase the number of usable alignments. This approach also implies that identifying streams originating from the same source must be done carefully. Different types of solar wind sources (e.g., coronal holes, streamers, and pseudo-streamers) exhibit distinct spatial extents and characteristic temporal evolution. When considering all wind types and source regions together, sufficiently strict criteria must be applied to footpoint location and stream departure time from the Sun, in order to guarantee nearly homogeneous source properties in space and time.  

\subsection{Backmapping Technique and Uncertainty Estimation}

We compute an extrapolation of the photospheric magnetic field using a potential field source surface (PFSS) model \citep{schatten1969}, with the source surface position fixed at $\rss = 2.5\,\rs$ \citep{arden2014}, where $\rs$ is the solar radius. The streamline tracing and PFSS modeling techniques used in this study are identical to those described in \cite{dakeyo2024b}. They are applied independently from PSP and Solar Orbiter, toward the Sun. The magnetic mapping is performed on \rev{30-minute} non-overlapping observation windows\rev{, and requires in-situ averaged plasma properties, including bulk velocity, plasma density, and magnetic field measurements}. Further details on the backmapping technique and the associated uncertainties relevant to source alignment are provided in Appendices~\ref{appendix:subsec_backmapping_technique} and~\ref{appendix:subsec_uncertainty_stream_assos}, respectively.

\subsection{Source Alignment and Plasma Identification Criteria}

\begin{figure*}[t!]
    \centering
    \includegraphics[width=18cm]{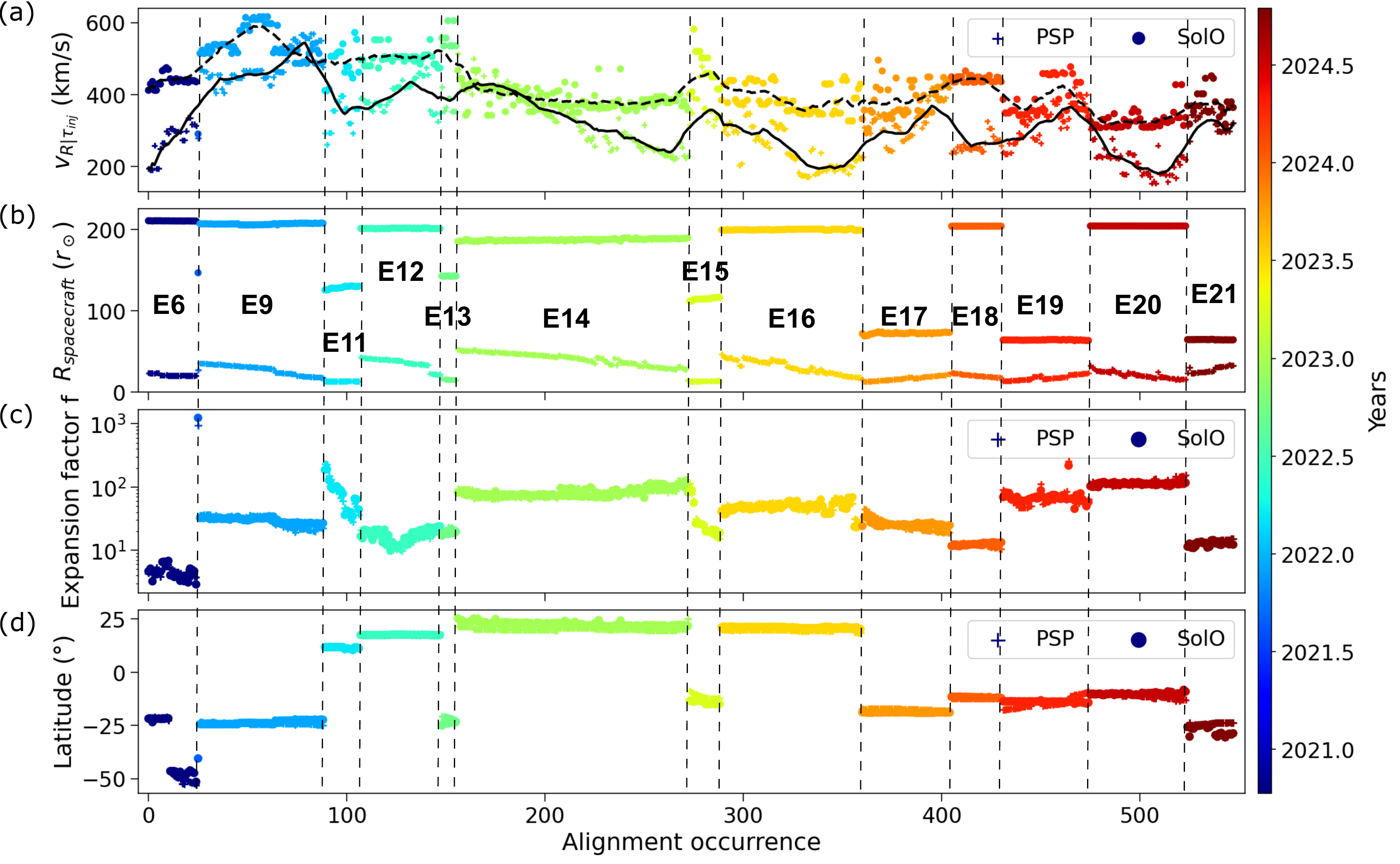}
    \caption{Source alignment occurrences between PSP and SO.
    Panel (a): Bulk solar wind speed;
    Panel (b): Spacecraft radial distance;
    Panel (c): Coronal magnetic field expansion factor derived from the PFSS model;
    Panel (d): Carrington latitude of the photospheric footpoints.
    The bulk speed in panel (a) is averaged over the injection time $\Tinj$ (see Section~\ref{appendix:sec_data_preprocessing} for details). Panels (a) and (b) show PSP and SO observations using horizontal crosses and dots, respectively.
    Mean trends are represented by running averages over 10 occurrences, shown with solid and dashed lines for PSP and SO, respectively, in panel (a).
    The corresponding PSP encounters for each alignment are indicated in panel (b) and separated by vertical dashed lines.
    }
    \label{fig_u_r_theta_fss}
\end{figure*}

The alignment criteria are divided into two categories: source identification criteria and in-situ identification criteria. Source identification is based on comparing magnetic properties at the stream footpoints, including the coronal expansion of magnetic flux tube (i.e. expansion factor), photospheric magnetic field polarity and intensity, as well as the spatial proximity of the footpoints from the two spacecraft.

The in-situ stream association criteria are based on the assumption that each stream follows a physically plausible solar wind radial evolution. These criteria include nearly interplanetary spherical expansion, polytropic temperature decrease, similar departure times of the streams observed at PSP and SO from the solar corona, comparable mass fluxes, and constrained speed evolution between the two probes. A complete description of all identification criteria is provided in Appendix~\ref{subsec:appendix_source_align_identif_crit}.

\rev{The identification criteria implicitly embed the following assumptions: (i) source regions are assumed to be spatially homogeneous and temporally stable over timescales of 2 days or more; (ii) close footpoint proximity is expected for similar plasma parcels; (iii) similar coronal expansion factors and photospheric magnetic fields are sufficient to uniquely identify a common source to streamlines; (iv) magnetic backmapping using a Parker spiral combined with a PFSS model (with fixed source surface at $2.5~\rs$) provides a reliable estimate of solar wind origin; (v) solar wind streams from a given source evolve in a broadly comparable way (e.g. approximately spherical expansion, polytropic temperature evolution, and near-conservation of mass flux); and (vi) turbulence properties and adaptive averaging scales can be meaningfully compared between spacecraft. 
}

\section{Dataset and Pre-processing}
\label{sec:data_preprocess}

\subsection{Data Collection}

The PSP data used in this study include partial proton moments from the SPAN-I (Solar Probe ANalyser) instrument of the SWEAP suite \citep{DOI_SPAN_A, Livi2022, Kasper2015SWEAP}, magnetic field measurements \citep{bale2016, DOI_MAG}, and electron density and temperature derived from quasi-thermal noise (QTN) observations \citep{QTN_ref_2020} from the FIELDS suite, as well as electron temperature measurements from the SPAN-E instrument \citep{SPAN_E_ref_2020}. Only measurements obtained below 0.25~au are retained, as this is the radial distance below which SPAN-I operates optimally, as detailed in the SPAN-I data release notes \citep{DOI_SPAN_A}.

Solar Orbiter observations consist of proton moments measured by the Proton and Alpha Sensor \citep[PAS;][]{owen_SWA2020} and magnetic field measurements from the MAG instrument \citep{Horbury_MAG2020}. 
\rev{Because $\Te$ is not yet directly measured at SO, its values are extrapolated from PSP measurements using a single polytropic index from \cite{dakeyo2022}.}
The magnetograms used for the PFSS extrapolation are the first realization of the ADAPT-GONG synoptic maps \citep{Arge2010}, with a temporal resolution of one map every 6 hours. Due to computational limitations, we \rev{only preform the mapping on one realization of} the ADAPT magnetogram. Nevertheless, we assume that changing the realization would only have a small effect on the proximity of the PSP and SO footpoints location, as in general the different realizations are similar to each other \citep{Li2021}.
\rev{For more details about the datasets, please refer to Appendix ~\ref{appendix:sec_data}.}

\renewcommand{\arraystretch}{1.2} 
\begin{table*}[t!]
\hspace{-3.2cm}
\scalebox{0.9}{
\begin{tabular}{|c|c|c|c|c|c|c|c|c|}
    \hline
     Date & Reference & PSP data & SO data & $<v>_{conj}$ (km/s) & Predicted & SA Detection & Known duration & SA Duration
     \tabularnewline
    \hline  
    27/09/2020 & \cite{Telloni2021} & \checkmark & \checkmark & 275 -- 439 & \checkmark & \checkmark & $\sim$ 2 h & $\sim$ 33 h
     \tabularnewline
    \hline 
    \multirow{2}{*}{29/04/2021}  & \cite{Berriot2024} & \multirow{2}{*}{\checkmark} & \multirow{2}{*}{\checkmark} & \multirow{2}{*}{-} & \multirow{2}{*}{\checkmark} & \multirow{2}{*}{-} & \multirow{2}{*}{$\sim$ 1.5 h} & \multirow{2}{*}{-} \\
    & \cite{Berriot2025} & & & & & & & 
     \tabularnewline
    \hline  
    11/08/2021 & - & \checkmark & \checkmark & 279 -- 291 & \checkmark & \checkmark & - & $\sim$ 0.5 h
     \tabularnewline
    \hline 
    18/09/2021 & - & - & - & - & \checkmark & - & - & -
     \tabularnewline
    \hline 
    19/11/2021 & - & \checkmark & \checkmark & 480 -- 541 & \checkmark & \checkmark & - & $\sim$ 47.5 h
     \tabularnewline
    \hline 
    \multirow{3}{*}{25/02/2022} & \cite{Rivera2024} & \multirow{3}{*}{\checkmark} & \multirow{3}{*}{\checkmark} & \multirow{3}{*}{342 -- 487} & \multirow{3}{*}{\checkmark} & \multirow{3}{*}{\checkmark} & \multirow{3}{*}{$\sim$ 9 h} & \multirow{3}{*}{$\sim$ 7 h}
    \\
    & \cite{Ervin2024a} & & & & & & & \\ 
    & \cite{Rivera2025} & & & & & & & 
    \tabularnewline 
    \hline
    06/04/2022 & - & - & \checkmark & - & \checkmark & - & - & -
     \tabularnewline
    \hline 
    31/05/2022 & - & \checkmark & \checkmark & 399 -- 502 & \checkmark & \checkmark & - & $\sim$ 57 h
     \tabularnewline
    \hline 
    06/09/2022 & - & \checkmark & \checkmark & 372 -- 558 & \checkmark & \checkmark & - & $\sim$ 6 h
     \tabularnewline
    \hline 
    10/12/2022 & \cite{Silwal2025} & \checkmark & \checkmark & 336 -- 389 & \checkmark & \checkmark & $\sim$ 1 h & $\sim$ 72.5 h 
     \tabularnewline
     \hline 
    17/03/2023 & \cite{Ervin2024b} & \checkmark & \checkmark & 366 -- 472 & \checkmark & \checkmark & - & $\sim$ 7.5 h
    \tabularnewline
    \hline
    17/04/2023 & - & - & \checkmark & - & \checkmark & - & - & -
     \tabularnewline
    \hline 
    21/06/2023 & - & \checkmark & \checkmark & 249 -- 388 & \checkmark & \checkmark & - & $\sim$ 68.5 h
     \tabularnewline
    \hline 
    28/09/2023 & - & \checkmark & \checkmark & 326 -- 387 & \checkmark & \checkmark & - & $\sim$ 24 h 
     \tabularnewline
    \hline 
    12/10/2023 & - & - & \checkmark & - & \checkmark & - & - & -
     \tabularnewline
    \hline 
    28/12/2023 & - & \checkmark & \checkmark & 261 -- 443 & \checkmark & \checkmark & - & $\sim$ 14 h
     \tabularnewline
    \hline 
    31/03/2024 & - & \checkmark & \checkmark & 324 -- 387 & \checkmark & \checkmark & - & $\sim$ 25.5 h
     \tabularnewline
    \hline 
    29/06/2024 & - & \checkmark & \checkmark & 210 -- 328 & \checkmark & \checkmark & - & $\sim$ 41.5 h
     \tabularnewline
    \hline 
    02/10/2024 & - & \checkmark & \checkmark & 325 -- 373 & - & \checkmark & - & $\sim$ 26 h
     \tabularnewline
    \hline 
\end{tabular}  
}
\caption{Comparison of the source alignment (SA) method detection capability with respect to existing radial and Parker spiral alignment studies.
First column: Date of the conjunction.
Second column: Reference to the corresponding study.
Third column: Availability of PSP data.
Fourth column: Availability of SO data.
Fifth column: Average wind speed over the entire SA conjunction interval, with the first and second values corresponding to PSP and SO, respectively.
Sixth column: predicted alignment.
Seventh column: Alignment detected by the SA method.
Eighth column: Known alignment duration from the study listed in the second column.
Ninth column: SA duration.
}
\label{tab:ref_other_studies_compar}
\end{table*}

\subsection{Observations Pre-processing}
\label{subsec:obs_pre_processing}

The high-sampling-rate datasets from PSP and SO are preprocessed using the same methodology. In the context of the backmapping algorithm, 30-minute time averages are used solely to determine the source regions of the in-situ solar wind observations. For the analysis of wind properties, plasma parameters are averaged over an adaptive time interval that depends on the radial distance $r$.   
As studied by \cite{Sioulas2025b}, the expansion of Alfvénic fluctuations with distance requires the observation window to be adapted to account for the stretching of the spatial and temporal scales. 
This approach also accounts for the fact that a given solar wind stream is stretched in the tangential direction during radial expansion \citep{Berriot2025}. It means that longer times are required to fully sample the stream at larger heliocentric distances. 
This aims to make more reliable the comparison between plasma and wave properties.
In addition, it provides a more consistent estimate of Alfvén wave properties at the injection scale \citep{Sioulas2025b}, which are assumed to play a key role in solar wind evolution \citep{Rivera2024, Rivera2025}.

We adopt a radial scaling of the injection frequency given by $\finj \propto r^{-1}$ \citep{Sioulas2023, Huang2025}, starting at the source surface above the super-expansion region, with a scale of $\rss = 2.5\:\rs$. The injection frequency is then defined as $\finj = 10^{-4} \times (r/r_{1au}) ^{-1}$~Hz 
, where $r_{1au}$ is the Earth -- Sun distance in solar radii. At 1~au, this scaling implies $\finj = 10^{-4}$~Hz, which is consistent with typical magnetic field power spectrum observations \citep{Alexandrova2013}. The corresponding adaptive averaging interval $\Tinj = 1/\finj$ is used to compute mean plasma properties.

These averaged quantities are synchronized with the 30-\rev{minute} mapping cadence (according to the closest observational time for both datasets), associating each stream source with its corresponding in-situ measurements averaged over the injection scale. Further details on the data pre-processing procedure are provided in Section~\ref{appendix:sec_data_preprocessing} of the Appendix.

\section{Stream Association Results}
\label{sec:stream_assoc_result}

\subsection{Source Alignment Occurrences}

After applying the full set of source alignment and stream association criteria based on the coronal and interplanetary properties of each stream, we present in Figure~\ref{fig_u_r_theta_fss} the source alignments identified between PSP and SO from 26 September 2020 to 3 October 2024. In total, we identify 548 non-overlapping source alignment streams, each lasting 30 minutes, corresponding to a total alignment duration of 16\,440 minutes. We recall that all quantities presented in the Sections~\ref{sec:stream_assoc_result} and \ref{sec:correl_results} are averaged over the turbulent injection time $\Tinj$.  

Panel~(a) shows the bulk solar wind speed measured at both spacecraft. It illustrates that, in most cases, solar wind streams undergo a continuous increase in speed within the inner heliosphere, between approximately $\sim$10~$\rs$ and $\sim$215~$\rs$. However, a small number of alignments exhibit weak acceleration or even a slight decrease in speed between the two probes (e.g., encounters E9 and E14).

The radial spacecraft separation shown in panel~(b) indicates that source alignments occur predominantly when SO is near 1~au, although alignments are not restricted to this configuration. Among the 16 PSP encounters analyzed, 13 include at least one source alignment interval of 30 minutes.

Regarding the coronal expansion factor and the photospheric footpoint latitude, we confirm that PSP and SO footpoints are in close proximity ($\leq 5 ^\circ$), as required by the source association criteria. Expansion factor values range between $\sim$5 and $\sim$1000, while footpoint latitudes span from approximately $-50^\circ$ to $+20^\circ$. These values are consistent with those reported in previous backmapping studies of coronal properties for both slow and fast solar wind streams \citep{dakeyo2024b}.

\begin{figure*}[t!]
    \centering
    \includegraphics[width=18cm]{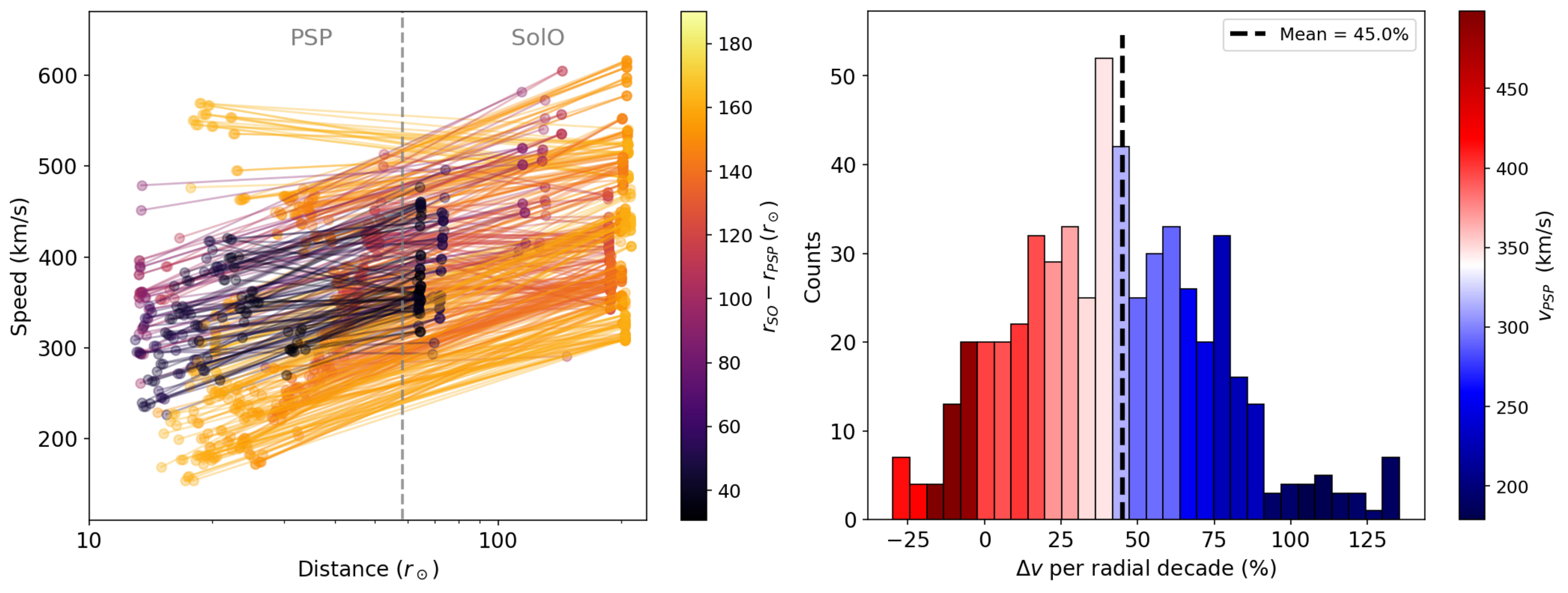}
    \caption{Bulk speed radial variation between PSP and SO for all source alignments. 
    Left panel: Bulk speed as a function of radial distance for source alignments.  
    The alignments are color-coded by the radial spacecraft separation between the two probes. 
    Right panel: Distribution of the relative solar wind speed increase $\Dv$ per radial decade, derived from PSP and SO source alignments. Each bin is color-coded by the average wind speed observed at PSP.  
    The vertical black dashed line indicates the average value.}
    \label{fig_spacecraft_separ_accel_align}
\end{figure*}

\subsection{PSP and SO Alignment Benchmark}

For comparison, we examine whether radial and Parker spiral alignments identified in previous studies (Table~\ref{tab:ref_other_studies_compar}) are also detected using the source alignment method. Among the five alignment conjunctions previously studied between PSP and SO \citep{Telloni2021, Berriot2024, Berriot2025, Rivera2024, Rivera2025, Ervin2024a, Ervin2024b, Silwal2025}, four are successfully detected by the source alignment approach.   

To further assess the reliability of the method, we compare our results with the predicted PSP--SO alignments described in \cite{Telloni2023} (also listed in Table~\ref{tab:ref_other_studies_compar}). 
These predictions are based on the theoretical orbit of the probe, with ballistic backmapping applied between PSP and SO as described in \cite{Telloni2021}. The predictions identify times when the probes are magnetically connected.
Of the 18 alignments predicted based on the spacecraft trajectories \citep{Telloni2023}, \rev{only 15 have data available at both spacecraft (affected by data gaps in either PSP or both PSP and SO observations). Within these 15 for which the backmapping procedure can be applied,} the source alignment method confirms 13 and identifies one additional alignment that was not originally predicted. The time interval over which plasma similarity is detected is generally longer for source alignment than for other alignment techniques, as expected given the longer temporal stability of solar wind sources.

\rev{Moreover, regarding the last missing alignment}, a mismatch of coronal properties (i.e. exceeding the stream association criteria) could cause the stream association to fail. To verify this, we relax the proximity in expansion factor criterion for the stream association from $3/4 < f_{\psp} / f_{\so} < 4/3$ originally, to $1/2< f_{\psp} / f_{\so} < 2$. This allows us to identify a source alignment conjunction lasting 30 minutes on the 28 April 2021, at 00:00, so a day before what is found in literature. \rev{As expected, we found that relaxing the criteria led to the detection of more source alignments}. However, this should be done carefully, bearing in mind the intended use of the stream association results.  

Overall, benchmarking the source alignment results against known and predicted alignments reported in the literature, demonstrates that our approach is consistent with existing methods.
Moreover, the strong agreement between source alignment detections and predicted alignments highlights the capability of the method to identify events analogous to single-stream radial \rev{and Parker spiral} evolution.

\subsection{Spacecraft Radial Separation}
\label{subsec:spacecraft_separ}

PSP and SO follow different heliocentric orbits and therefore observe different phases of solar wind evolution. Previous studies have shown that the mean solar wind speed beyond 0.3~au follows a logarithmic dependence on radial distance, $v = a \log r + b$, where $a$ and $b$ are constants determined by fitting multi-spacecraft observations \citep{maksimovic2020, dakeyo2022, Halekas2023, dakeyo2024a}. To enable comparison between alignments observed at different radial separations, we rescale the observed velocity difference $v_\so - v_\psp$ to a reference radial decade, corresponding to $r_\so = 10 \times r_\psp$.

This is achieved by dividing the observed velocity difference by a spacecraft separation  factor $\Css$ such as~:
\begin{align}
    \Dv = \frac{v_\so - v_\psp}{\Css} 
    \quad
    \text{with} \quad
    \Css =  \log \bigg( \frac{\rsolo}{\rpsp}  \bigg) \ / \ \log 10.
    \label{eq:Delta_v_express}
\end{align}
Because the PSP--SO separation is typically close to one radial decade, this correction remains modest and primarily ensures that all alignment pairs provide a comparable measure of $\Dv$.

Panel~(a) of Figure~\ref{fig_spacecraft_separ_accel_align} compares the wind speeds measured at PSP and SO. We find that the solar wind undergoes a significant average speed increase of +45\% ($+$147~km/s) per radial decade (e.g., between 15~$\rs$ and 150~$\rs$), for an average PSP wind speed of $\sim$328~km/s.  
A strong anticorrelation is present between $\Dv$ and $v_\psp$ (panel~(b) of Figure~\ref{fig_spacecraft_separ_accel_align}), whereby slower wind speeds at PSP are associated with larger speed increases, with a Pearson correlation coefficient of $-0.77$. This trend is consistent with previous studies that analyzed mixed solar wind streams at different radial distances without explicit alignment classification \citep{maksimovic2020, dakeyo2022, Halekas2023}.

\begin{figure*}[t]
    \centering
    \includegraphics[width=18cm]{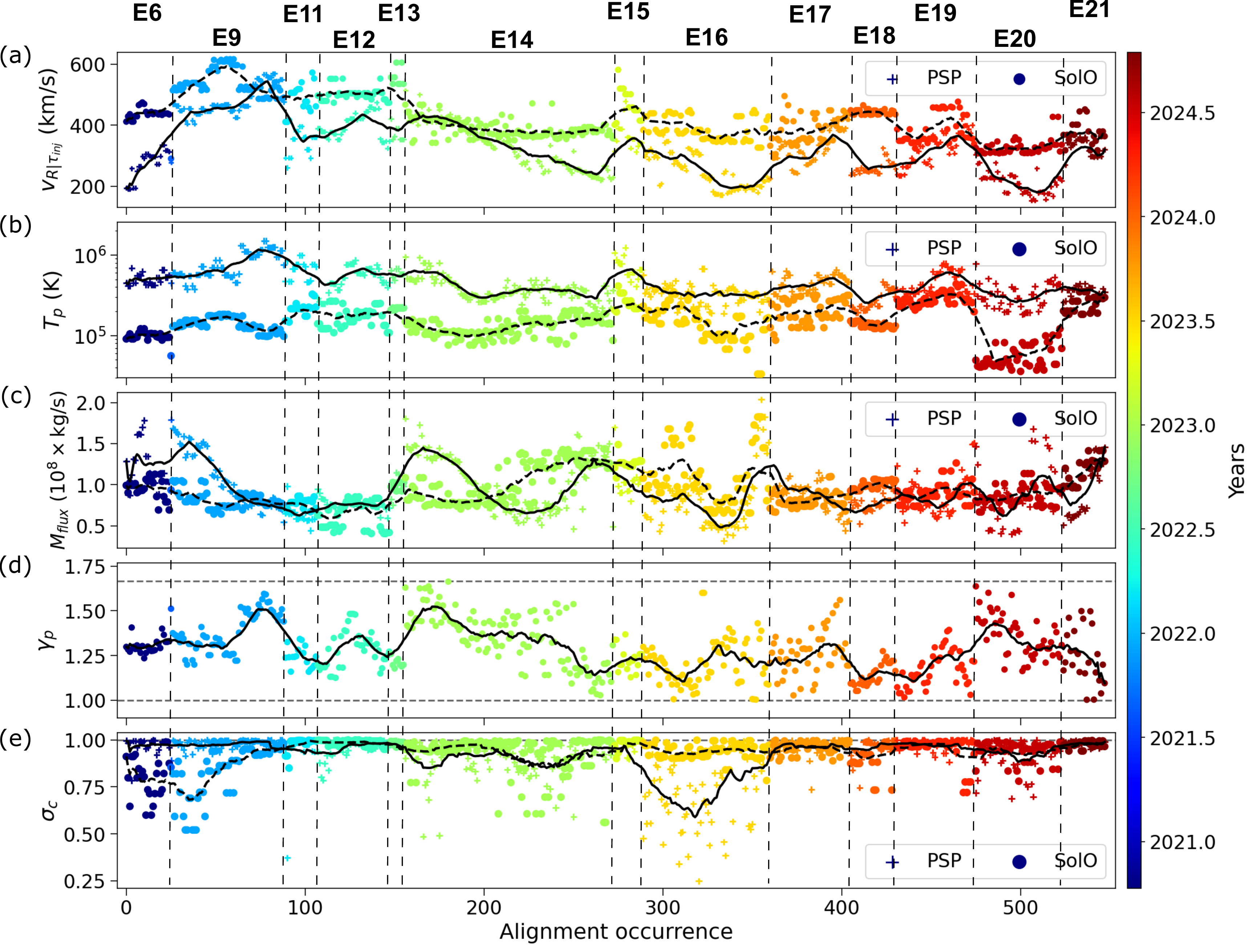}
    \caption{Same format as Figure~\ref{fig_u_r_theta_fss} for additional plasma parameters.
    Panel (a): bulk solar wind speed;
    Panel (b): proton temperature;
    Panel (c): Mass flux defined by $M_{flux} = (m_p + m_e)  \ v_{R|\Tinj} \ n \ r^2 $;
    Panel (d): proton polytropic index, derived from the logarithmic slope of the polytropic relation between $\Tp$ and $n$ between PSP and SO for each alignment.
    All quantities are averaged or computed over the injection time $\Tinj$ (see Section~\ref{appendix:sec_data_preprocessing} for details). PSP and SO observations are shown as crosses and dots, respectively, in panels (a)--(d). Mean trends are indicated by running averages over 10 occurrences, with solid and dashed lines for PSP and SO, respectively, and a solid line in panel (d). The corresponding PSP encounters are indicated at the top of the figure and separated by vertical dashed lines.}
    \label{fig_u_Tp_np_gammap_sigma_c}
\end{figure*}

\subsection{Plasma Parameters Radial Evolution}

Figure~\ref{fig_u_Tp_np_gammap_sigma_c} presents the radial evolution of plasma parameters averaged over $\Tinj$ at PSP and SO. Panel~(a) is identical to that in Figure~\ref{fig_u_r_theta_fss} and is shown for reference to highlight any dependence on wind speed.

The proton temperatures shown in panel~(b) display a clear correlation with the corresponding bulk solar wind speeds observed at PSP and SO. The magnitude of the temperature decrease between the two spacecraft varies from stream to stream. 

The mass flux exhibits comparable values at PSP and SO, as defined by the stream association criterion defined in Appendix~\ref{subsec:appendix_source_align_identif_crit}, although deviations from mass flux conservation are apparent. Enhanced mass flux at SO may indicate compression, whereas reduced values may reflect rarefaction. These last could potentially be caused by stream interactions. 

The proton polytropic index, derived from the logarithmic fit between $\Tp$ and $n$, is shown in panel~(d). The index lies between the isothermal ($\gamp = 1$) and adiabatic ($\gamp = 5/3$) limits, statistically confirming that protons are heated along individual solar wind streams in the interplanetary medium.

The cross helicity defined as $\sigma_c = (2 \ \dvvect . \dbvect) / (\dv^2 + \db^2)$,  
where $\dvvect$ and $\dbvect$ are respectively the velocity and magnetic field fluctuation vector computed over $\Tinj$ (see Appendix~\ref{appendix:subsec_fluctuation_comput}), quantifies the degree of Alfvénicity, that is the correlation between $\dvvect$ and $\dbvect$.
We can see from the values of $\sigma_c$ shown on the panel~(e), that the solar wind observed within source alignments is predominantly Alfvénic, although some slow wind streams exhibit a reduction in Alfvénicity. In terms of solar cycle dependence, alignments are more frequently observed during solar maximum conditions (after approximately 03/2022, see color scale) than during solar minimum.  
This may partly explain the generally high levels of Alfvénicity observed for many source alignments \citep{Damicis2021}.

\begin{figure*}[t!]
    \centering
    \includegraphics[width=18cm]{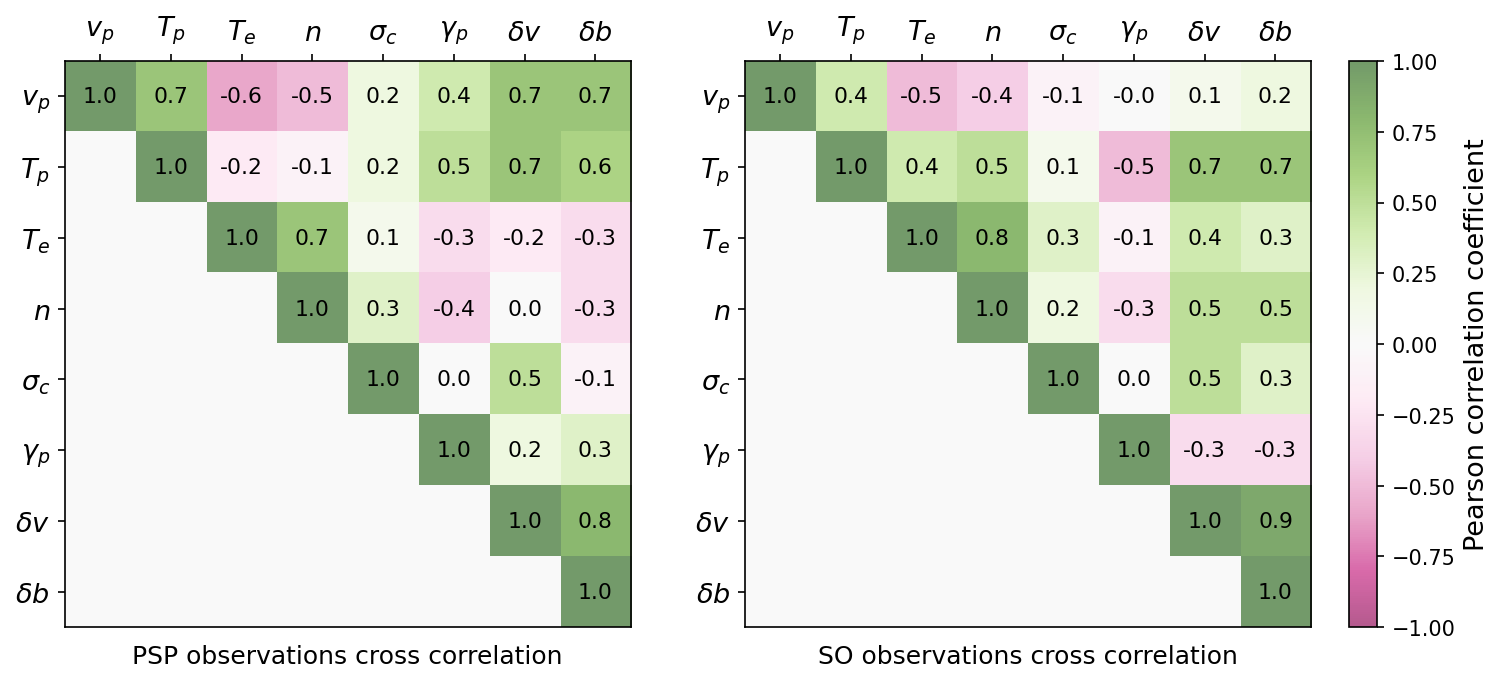}
    \caption{Correlation matrices of plasma parameters observed at PSP and SO within source alignments. Left panel: cross-correlation matrix for PSP alignment observations. Right panel: cross-correlation matrix for SO alignment observations. All quantities are averaged  over $\Tinj$. Because the quantities span different orders of magnitude and are typically related by power laws, logarithmic values are used when computing correlations. As the diagonal elements represent autocorrelations, their coefficients are equal to unity by definition. Owing to the symmetry of the matrices, only the upper-right portions are displayed.}
    \label{fig_correl_var_psp_solo_indep}
\end{figure*}

\section{Correlation of the Wind Characteristics within the Source Alignment}
\label{sec:correl_results}

\subsection{Cross-Correlation of the Plasma Parameters at PSP and SO}

To further characterize the differences between source-aligned and statistical approaches, we investigate how the observed plasma properties depend on the average stream bulk speed and how they correlate with one another. To this end, we compute cross-correlation matrices for a set of solar wind characteristics independently at PSP and SO (Figure~\ref{fig_correl_var_psp_solo_indep}). Throughout this section, correlations are quantified using the Pearson correlation coefficient, which ranges from $-1$ to $+1$, corresponding to perfect anticorrelation and perfect correlation, respectively.

Consistent with well-established solar wind relationships, we observe that $v_p$ at both PSP and SO is correlated with $\Tp$ (with coefficients of +0.7 and +0.4, respectively) and anticorrelated with $n$ (-0.5 and -0.4) and $\Te$ (-0.6 and -0.5). 
\rev{We remind that due to the lack of electron temperature at SO, $\Te$ are extrapolated from PSP observations, as mentioned in Section~\ref{subsec:obs_pre_processing}.}
Consequently, correlations involving $\Te$ at SO should be interpreted with caution.

For other plasma parameters, several correlations evolve with radial distance. For instance, $\Tp$ is correlated with $\gamp$ at PSP (+0.5) but becomes anticorrelated at SO (-0.5). In addition, the correlations of $v_p$ with $\Tp$, $\dv$, and $\db$ decrease from (+0.7, +0.7, +0.7) at PSP to (+0.4, +0.1, +0.2) at SO, indicating that the radial expansion can obscure the relationship in between solar wind parameters.

\subsection{Cross-Correlation of the Plasma Parameters Radial Evolution}

To account for the observed increase in bulk velocity, the in-situ plasma properties are expected to evolve between PSP and SO. While correlations among bulk plasma parameters at a given distance have been extensively studied, correlations between the radial evolution of the parameters along individual streams, are studied less. The speed increase $\Dv$ between PSP and SO is expected to arise from pressure gradients associated with decreases in proton temperature ($\DTp$), electron temperature ($\DTe$), and density ($\Dnp$). In addition, Alfvénic fluctuations are expected to correlate with proton heating and may also contribute to acceleration through wave pressure gradients.

\begin{figure}[t]
    \centering
    \includegraphics[width=8.5cm]{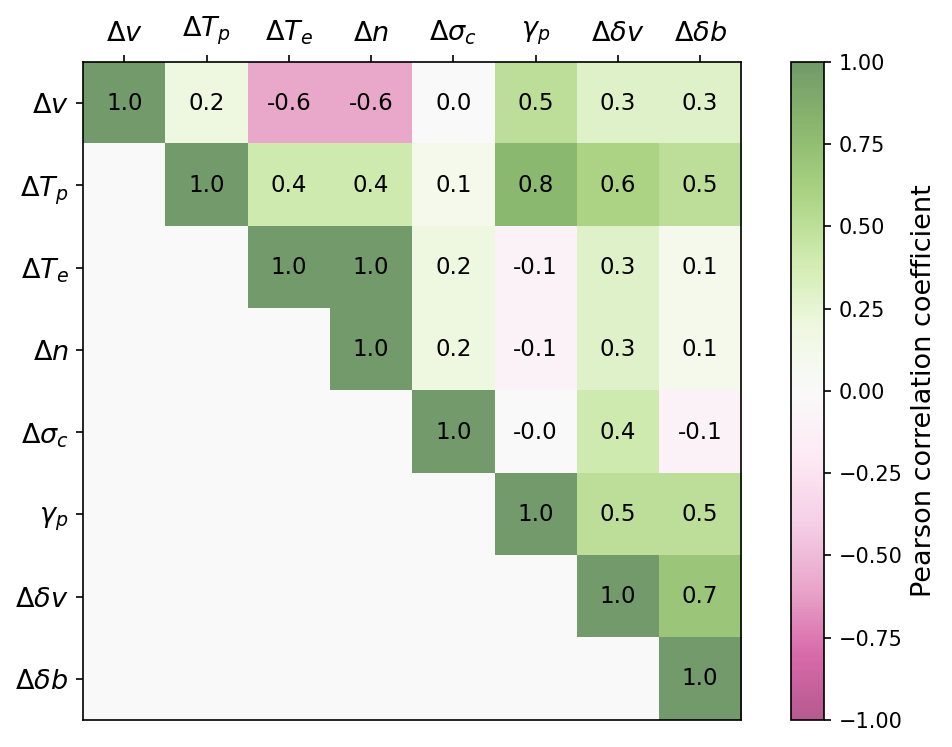}
    \caption{Same format as Figure~\ref{fig_correl_var_psp_solo_indep}, but applied to the radial variations of plasma parameters within PSP and SO source alignments. 
    Radial variations are computed as $\Delta X = X_\psp - X_\so$ for a given observable $X$. The polytropic index $\gamp$, derived from two-point measurements, is compared directly with other varying observables. Because the quantities span different orders of magnitude and are typically related by power laws, logarithmic values \rev{of $X$ at PSP and SO} are used when computing $\Delta X$ (except for $X=\sigma_c$). All quantities are averaged or computed over $\Tinj$. 
    }
    \label{fig_cross_correl_dvar_psp_solo}
\end{figure}

Figure~\ref{fig_cross_correl_dvar_psp_solo} presents the cross-correlation matrix of the radial variations of the different plasma parameters. We observe that $\Dv$ anticorrelates with $\Dnp$ and correlates with $\gamp$ (with coefficients of $-0.6$ and $+0.5$, respectively). This behavior supports the idea that stronger density decreases (i.e. pressure gradients) are associated with greater increases in speed. This is also consistent with the principle of mass flux conservation.

We also find that $\Dv$ anticorrelates with $\DTe$ (-0.6), indicating that larger speed increases are associated with stronger decreases in electron temperature. We recall that $\Te$ is directly measured only at PSP and extrapolated at SO using an average $\game$ value from \cite{dakeyo2022}. As a result, correlations involving $\DTe$ reflect average trends based on this assumed $\game$ and do not capture event-to-event variability in electron temperature evolution. \rev{Including such a variability would lead to smaller correlations between $\DTe$ and $\Dv$. Thus, the resulting value of -0.6 should be taken as an upper bound for what is observed.}

In contrast, correlations between $\Dv$ and $\Ddv$, $\Ddb$, or $\Dsigmac$ are relatively weak (all $\lesssim 0.3$ in absolute value). This suggests that wave pressure gradients associated with \rev{a decrease in fluctuations}, as well as changes in Alfvénicity, \rev{does not appear} to be the dominant drivers of solar wind acceleration in the inner heliosphere.

Finally, we observe positive correlations between $\DTp$ and both $\Ddv$ and $\Ddb$ (+0.6 and +0.5, respectively). While a relationship between proton heating and wave dissipation is expected, the positive sign of these correlations is somewhat surprising, as a decrease in fluctuations might instead be expected to correspond to slower proton temperature decreases.
This apparent discrepancy indicates the limitations of the present correlation approach. We should \rev{note} that the number of physical processes \rev{leading to the radial evolution of} $\Tp$, $\dv$ and $\db$ \rev{could potentially be} the reason. The spherical expansion primarily induces a radial variation caused by adiabatic expansion for $\DTp$ \citep{parker1958}, and by the conservation of wave action for $\Ddv$ and $\Ddb$ \citep{Alazraki1971, belcher1971}. Both physical processes are accelerating the solar wind. However, heating processes and potential stream interactions (SIRs) are also occurring, resulting in additional radial scaling beyond that induced by the spherical expansion. 

Therefore, it is not possible to provide more detailed interpretations of the radial evolution of plasma and wave parameters without disentangling the different causes of the radial variation (spherical work, thermal exchanges by heating and SIRs). As this task is complex and requires the integration of theoretical plasma and wave dynamics and observations, we let this to be addressed in future work. 

\section{Conclusion}
\label{sec:conclusion}

This paper presents the first results obtained using a newly developed source alignment method applied to Parker Solar Probe (PSP) and Solar Orbiter (SO) observations. The method identifies solar wind streams observed by the two spacecraft that originate from the same source region in the solar photosphere. Time propagation from the source is taken into account in order to determine when a given stream leaves the solar atmosphere and is subsequently observed by PSP and SO.

\rev{To guarantee the reliability of the stream association, we set source alignment criteria based on the coronal and interplanetary properties of each stream.
Criteria embed the following assumptions~: (i) source regions are mostly spatially homogeneous and temporally stable over timescales of at least 2 days; (ii) close footpoint proximity is expected for similar plasma parcels originating the same source, (iii) similar coronal expansion factor and photospheric magnetic field magnitude are sufficient to uniquely identify a common source to streamlines; (iv) magnetic backmapping using a Parker spiral combined with a PFSS model (with fixed source surface at $2.5~\rs$) mostly provides a reliable estimate of solar wind origin; (v) solar wind streams from a given source evolve in a broadly comparable way (e.g. approximately spherical expansion, polytropic temperature evolution, and near-conservation of mass flux); and (vi) to improve the data comparison from different spacecraft due to the plasma expansion, an adaptive averaging timescale is applied to meaningfully compare turbulence and wind bulk properties data between spacecraft. However, while all data between spacecraft relies on the adaptive time interval, for numerical convenience, and exclusively for the backmapping computation, footpoints are estimated every 30 minutes instead.
After applying all criteria and the appropriate data averaging for the radial evolution analysis, } 
we identified 548 non-overlapping 30-minute source alignment streams, totaling 16\,440 minutes of alignments. 

Among the 18 \rev{alignment intervals (i.e. time interval comprising several stream alignments)} predicted between PSP and SO based on their \rev{planned} trajectories
\rev{\citep{Telloni2023}, only 15 of them can be tested since it requires data available at both spacecraft. Among these 15 testable predictions, 13} are \rev{observationally} detected using the source alignment method. \rev{One additional unpredicted alignment interval is as well detected, which raises the total to 14 detected source alignment intervals.}

Source alignments are found to be significantly more frequent than with previously used alignment methods (radial and Parker spiral alignments), making it possible to build the first \rev{comprehensive} statistics on the radial evolution of individual solar wind streams.

Having established the reliability of the method, we conducted a statistical investigation of the radial evolution of single-stream evolution-like events. We observe a significant increase in the average solar wind speed of $+$147~km/s (+45\%) per radial decade (e.g. between 15~$\rs$ and 150~$\rs$), for an average wind speed of $\sim$328~km/s at PSP. We also find that slower wind speeds at PSP correspond to larger speed increases (Pearson correlation coefficient of $-0.77$). These trends are consistent with previous statistical studies of solar wind radial evolution that do not rely on aligned streams \citep{maksimovic2020, dakeyo2022}.
This demonstrates that above 10 solar radii, the wind velocity cannot be considered constant, as it has not yet reached an asymptotic state. This supports the importance of considering an accelerating solar wind profile when studying the interaction of \rev{different solar winds (within stream interaction regions)}, creating prediction models (i.e. space weather forecasts), or studying long distance radial evolution of the solar wind. 

An analysis of the cross-correlations between plasma parameters and their radial variations between PSP and SO shows that the speed increase ($\Dv$) is primarily associated with a decrease in plasma density ($\Dnp$) and electron temperature ($\DTe$), with an anticorrelation coefficient of $-0.6$ for both. In contrast, $\Dv$ exhibits only weak correlations with the decreases in bulk velocity and Alfvén velocity fluctuation ($\Ddv$ and $\Ddb$ respectively) and proton temperature ($\DTp$). 
These results support earlier findings suggesting that electron thermal pressure plays \rev{a non negligible} role in solar wind acceleration, above the main acceleration region $\gtrsim 15~\rs$ \citep{halekas2022, dakeyo2022}. \rev{However, $\DTe$ strongly depends on the assumed scaling of $\Te$ from \citet{dakeyo2022} due to the lack of $\Te$ observations at SO. Thus, the present statements regarding $\DTe$ should be interpreted with caution.} 

We observe that the variation in $\DTp$ exhibits a positive correlation coefficient with both $\Ddv$ and $\Ddb$ \rev{(a proxy for Alfvénic fluctuations)}, rather than the negative correlation coefficients expected from theory \citep{verdini2007, chandran2009, reville2020, Sioulas2025b}. The causes of the radial variation of all these quantities are numerous and complex, making it difficult to interpret any cross-correlation results (e.g. spherical work, heating processes, stream interactions). These results highlight the need for a more detailed comparison of observational alignments with theoretical models. A complete energy budget, for example, would help us to disentangle the roles of the different physical processes, and improve our understanding of the radial energy transfer and acceleration processes in the solar wind.

\acknowledgments{We acknowledge the NASA Parker Solar Probe Mission and the SWEAP team led by J. Kasper for use of data. We thank the instrumental Solar Wind Analyser team (SWA) for valuable discussions. This research was funded by the European Research Council ERC SLOW\_SOURCE (DLV-819189) project. This work was supported by CNRS Occitanie Ouest and LIRA. We recognise the collaborative and open nature of knowledge creation and dissemination, under the control of the academic community as expressed by Camille Noûs at http://www.cogitamus.fr/indexen.html. The work at AIP carried out by A.P.R. was supported by the Alexander von Humboldt foundation.
TE acknowledges funding from The Chuck Lorre Family Foundation Big Bang Theory Graduate Fellowship and NASA contract NNN06AA01C. Mingzhe Liu also acknowledges partial support from NASA HGIO grant 80NSSC25K7689.
}

\bibliography{bibliography}{}

@misc{DOI_SPAN_A,
  doi = {10.48322/YPYH-S325},
  url = {https://hpde.io/NASA/NumericalData/ParkerSolarProbe/SWEAP/SPAN-A/Level3/ProtonPartialMoments/InstrumentFrame/PT7S},
  author = {Livi,  Roberto and Larson,  Davin E. and Rahmati,  Ali},
  title = {PSP Solar Wind Electrons Alphas and Protons (SWEAP) SPAN-A Proton Distribution Function,  Partial Moments,  Instrument Frame,  Level 3 (L3),  7 s Data},
  publisher = {NASA Space Physics Data Facility},
  year = {2020},
  copyright = {Creative Commons Zero v1.0 Universal}
}

@misc{DOI_MAG,
  doi = {10.48322/0YY0-BA92},
  url = {https://hpde.io/NASA/NumericalData/ParkerSolarProbe/FIELDS/MAG/Level2/RTN/FullResolution/PT0.003413S},
  author = {Bale,  Stuart D. and MacDowall,  Robert J. and Koval,  Andriy and Pulupa,  Marc and Quinn,  Timothy and Schroeder,  Peter},
  title = {PSP FIELDS Fluxgate Magnetometer (MAG) Magnetic Field Vectors,  Radial-Tangential-Normal,  RTN,  Coordinates,  Full Resolution,  Level 2 (L2),  3.413 ms Data},
  publisher = {NASA Space Physics Data Facility},
  year = {2020},
  copyright = {Creative Commons Zero v1.0 Universal}
}

@ARTICLE{bale2016,
       author = {{Bale}, S.~D. and {Goetz}, K. and {Harvey}, P.~R. and {Turin}, P. and {Bonnell}, J.~W. and {Dudok de Wit}, T. and {Ergun}, R.~E. and {MacDowall}, R.~J. and {Pulupa}, M. and {Andre}, M. and {Bolton}, M. and {Bougeret}, J. -L. and {Bowen}, T.~A. and {Burgess}, D. and {Cattell}, C.~A. and {Chandran}, B.~D.~G. and {Chaston}, C.~C. and {Chen}, C.~H.~K. and {Choi}, M.~K. and {Connerney}, J.~E. and {Cranmer}, S. and {Diaz-Aguado}, M. and {Donakowski}, W. and {Drake}, J.~F. and {Farrell}, W.~M. and {Fergeau}, P. and {Fermin}, J. and {Fischer}, J. and {Fox}, N. and {Glaser}, D. and {Goldstein}, M. and {Gordon}, D. and {Hanson}, E. and {Harris}, S.~E. and {Hayes}, L.~M. and {Hinze}, J.~J. and {Hollweg}, J.~V. and {Horbury}, T.~S. and {Howard}, R.~A. and {Hoxie}, V. and {Jannet}, G. and {Karlsson}, M. and {Kasper}, J.~C. and {Kellogg}, P.~J. and {Kien}, M. and {Klimchuk}, J.~A. and {Krasnoselskikh}, V.~V. and {Krucker}, S. and {Lynch}, J.~J. and {Maksimovic}, M. and {Malaspina}, D.~M. and {Marker}, S. and {Martin}, P. and {Martinez-Oliveros}, J. and {McCauley}, J. and {McComas}, D.~J. and {McDonald}, T. and {Meyer-Vernet}, N. and {Moncuquet}, M. and {Monson}, S.~J. and {Mozer}, F.~S. and {Murphy}, S.~D. and {Odom}, J. and {Oliverson}, R. and {Olson}, J. and {Parker}, E.~N. and {Pankow}, D. and {Phan}, T. and {Quataert}, E. and {Quinn}, T. and {Ruplin}, S.~W. and {Salem}, C. and {Seitz}, D. and {Sheppard}, D.~A. and {Siy}, A. and {Stevens}, K. and {Summers}, D. and {Szabo}, A. and {Timofeeva}, M. and {Vaivads}, A. and {Velli}, M. and {Yehle}, A. and {Werthimer}, D. and {Wygant}, J.~R.},
        title = "{The FIELDS Instrument Suite for Solar Probe Plus. Measuring the Coronal Plasma and Magnetic Field, Plasma Waves and Turbulence, and Radio Signatures of Solar Transients}",
      journal = {\ssr},
         year = 2016,
        month = dec,
       volume = {204},
       number = {1-4},
        pages = {49-82},
          doi = {10.1007/s11214-016-0244-5},
       adsurl = {https://ui.adsabs.harvard.edu/abs/2016SSRv..204...49B}
}

@INPROCEEDINGS{1981_ref_helios,
       author = {{Porsche}, H.},
        title = "{HELIOS mission: Mission objectives, mission verification, selected results}",
    booktitle = {Solar System and its Exploration},
         year = 1981,
       editor = {{Burke}, W.~R.},
       series = {ESA Special Publication},
       volume = {164},
        month = nov,
        pages = {43-50},
       adsurl = {https://ui.adsabs.harvard.edu/abs/1981ESASP.164...43P}
}

@article{fox2016solar,
       author = {{Fox}, N.~J. and {Velli}, M.~C. and {Bale}, S.~D. and {Decker}, R. and {Driesman}, A. and {Howard}, R.~A. and {Kasper}, J.~C. and {Kinnison}, J. and {Kusterer}, M. and {Lario}, D. and {Lockwood}, M.~K. and {McComas}, D.~J. and {Raouafi}, N.~E. and {Szabo}, A.},
        title = "{The Solar Probe Plus Mission: Humanity's First Visit to Our Star}",
      journal = {\ssr},
         year = 2016,
        month = dec,
       volume = {204},
       number = {1-4},
        pages = {7-48},
          doi = {10.1007/s11214-015-0211-6},
       adsurl = {https://ui.adsabs.harvard.edu/abs/2016SSRv..204....7F}
}

@article{Lopez1986solar,
       author = {{Lopez}, R.~E. and {Freeman}, J.~W.},
        title = "{Solar wind proton temperature-velocity relationship}",
      journal = {\jgr},
         year = 1986,
        month = feb,
       volume = {91},
       number = {A2},
        pages = {1701-1705},
          doi = {10.1029/JA091iA02p01701},
       adsurl = {https://ui.adsabs.harvard.edu/abs/1986JGR....91.1701L}
}

@article{Elliott2012temporal,
       author = {{Elliott}, H.~A. and {Henney}, C.~J. and {McComas}, D.~J. and {Smith}, C.~W. and {Vasquez}, B.~J.},
        title = "{Temporal and radial variation of the solar wind temperature-speed relationship}",
      journal = {Journal of Geophysical Research (Space Physics)},
         year = 2012,
        month = sep,
       volume = {117},
       number = {A9},
          eid = {A09102},
        pages = {A09102},
          doi = {10.1029/2011JA017125},
       adsurl = {https://ui.adsabs.harvard.edu/abs/2012JGRA..117.9102E}
}

@ARTICLE{maksimovic2020,
       author = {{Maksimovic}, M. and {Bale}, S.~D. and {Ber{\v{c}}i{\v{c}}}, L. and {Bonnell}, J.~W. and {Case}, A.~W. and {Dudok de Wit}, T. and {Goetz}, K. and {Halekas}, J.~S. and {Harvey}, P.~R. and {Issautier}, K. and {Kasper}, J.~C. and {Korreck}, K.~E. and {Jagarlamudi}, V. Krishna and {Lahmiti}, N. and {Larson}, D.~E. and {Lecacheux}, A. and {Livi}, R. and {MacDowall}, R.~J. and {Malaspina}, D.~M. and {Martinovi{\'c}}, M.~M. and {Meyer-Vernet}, N. and {Moncuquet}, M. and {Pulupa}, M. and {Salem}, C. and {Stevens}, M.~L. and {{\v{S}}tver{\'a}k}, {\v{S}}. and {Velli}, M. and {Whittlesey}, P.~L.},
        title = "{Anticorrelation between the Bulk Speed and the Electron Temperature in the Pristine Solar Wind: First Results from the Parker Solar Probe and Comparison with Helios}",
      journal = {\apjs},
         year = 2020,
        month = feb,
       volume = {246},
       number = {2},
          eid = {62},
        pages = {62},
          doi = {10.3847/1538-4365/ab61fc},
       adsurl = {https://ui.adsabs.harvard.edu/abs/2020ApJS..246...62M}
}

@ARTICLE{Demoulin2009T-Vcorrelation,
       author = {{D{\'e}moulin}, P.},
        title = "{Why Do Temperature and Velocity Have Different Relationships in the Solar Wind and in Interplanetary Coronal Mass Ejections?}",
      journal = {\solphys},
         year = 2009,
        month = jun,
       volume = {257},
       number = {1},
        pages = {169-184},
          doi = {10.1007/s11207-009-9338-5},
       adsurl = {https://ui.adsabs.harvard.edu/abs/2009SoPh..257..169D}
}

@article{Kasper2015SWEAP,
      author = {{Kasper}, Justin C. and {Abiad}, Robert and {Austin}, Gerry and {Balat-Pichelin}, Marianne and {Bale}, Stuart D. and {Belcher}, John W. and {Berg}, Peter and {Bergner}, Henry and {Berthomier}, Matthieu and {Bookbinder}, Jay and {Brodu}, Etienne and {Caldwell}, David and {Case}, Anthony W. and {Chandran}, Benjamin D.~G. and {Cheimets}, Peter and {Cirtain}, Jonathan W. and {Cranmer}, Steven R. and {Curtis}, David W. and {Daigneau}, Peter and {Dalton}, Greg and {Dasgupta}, Brahmananda and {DeTomaso}, David and {Diaz-Aguado}, Millan and {Djordjevic}, Blagoje and {Donaskowski}, Bill and {Effinger}, Michael and {Florinski}, Vladimir and {Fox}, Nichola and {Freeman}, Mark and {Gallagher}, Dennis and {Gary}, S. Peter and {Gauron}, Tom and {Gates}, Richard and {Goldstein}, Melvin and {Golub}, Leon and {Gordon}, Dorothy A. and {Gurnee}, Reid and {Guth}, Giora and {Halekas}, Jasper and {Hatch}, Ken and {Heerikuisen}, Jacob and {Ho}, George and {Hu}, Qiang and {Johnson}, Greg and {Jordan}, Steven P. and {Korreck}, Kelly E. and {Larson}, Davin and {Lazarus}, Alan J. and {Li}, Gang and {Livi}, Roberto and {Ludlam}, Michael and {Maksimovic}, Milan and {McFadden}, James P. and {Marchant}, William and {Maruca}, Bennet A. and {McComas}, David J. and {Messina}, Luciana and {Mercer}, Tony and {Park}, Sang and {Peddie}, Andrew M. and {Pogorelov}, Nikolai and {Reinhart}, Matthew J. and {Richardson}, John D. and {Robinson}, Miles and {Rosen}, Irene and {Skoug}, Ruth M. and {Slagle}, Amanda and {Steinberg}, John T. and {Stevens}, Michael L. and {Szabo}, Adam and {Taylor}, Ellen R. and {Tiu}, Chris and {Turin}, Paul and {Velli}, Marco and {Webb}, Gary and {Whittlesey}, Phyllis and {Wright}, Ken and {Wu}, S.~T. and {Zank}, Gary},
        title = "{Solar Wind Electrons Alphas and Protons (SWEAP) Investigation: Design of the Solar Wind and Coronal Plasma Instrument Suite for Solar Probe Plus}",
      journal = {\ssr},
         year = 2016,
        month = dec,
       volume = {204},
       number = {1-4},
        pages = {131-186},
          doi = {10.1007/s11214-015-0206-3},
       adsurl = {https://ui.adsabs.harvard.edu/abs/2016SSRv..204..131K}
}

@ARTICLE{dakeyo2022,
       author = {{Dakeyo}, Jean-Baptiste and {Maksimovic}, Milan and {D{\'e}moulin}, Pascal and {Halekas}, Jasper and {Stevens}, Michael L.},
        title = "{Statistical Analysis of the Radial Evolution of the Solar Winds between 0.1 and 1 au and Their Semiempirical Isopoly Fluid Modeling}",
      journal = {\apj},
         year = 2022,
        month = dec,
       volume = {940},
       number = {2},
          eid = {130},
        pages = {130},
          doi = {10.3847/1538-4357/ac9b14},
archivePrefix = {arXiv},
       eprint = {2207.03898},
 primaryClass = {astro-ph.SR},
       adsurl = {https://ui.adsabs.harvard.edu/abs/2022ApJ...940..130D}
}

@ARTICLE{dakeyo2024a,
       author = {{Dakeyo}, J. -B. and {Badman}, S.~T. and {Rouillard}, A.~P. and {R{\'e}ville}, V. and {Verscharen}, D. and {D{\'e}moulin}, P. and {Maksimovic}, M.},
        title = "{Radial evolution of the accuracy of ballistic solar wind backmapping}",
      journal = {\aap},
         year = 2024,
        month = jun,
       volume = {686},
          eid = {A12},
        pages = {A12},
          doi = {10.1051/0004-6361/202348892},
       adsurl = {https://ui.adsabs.harvard.edu/abs/2024A&A...686A..12D}
}

@ARTICLE{dakeyo2024b,
       author = {{Dakeyo}, J-B. and {Rouillard}, A.~P. and {R{\'e}ville}, V. and {D{\'e}moulin}, P. and {Maksimovic}, M. and {Chapiron}, A. and {Pinto}, R.~F. and {Louarn}, P.},
        title = "{Testing the flux tube expansion factor -- solar wind speed relation with Solar Orbiter data}",
      journal = {arXiv e-prints},
         year = 2024,
        month = aug,
          eid = {arXiv:2408.06155},
        pages = {arXiv:2408.06155},
          doi = {10.48550/arXiv.2408.06155},
archivePrefix = {arXiv},
       eprint = {2408.06155},
 primaryClass = {astro-ph.SR},
       adsurl = {https://ui.adsabs.harvard.edu/abs/2024arXiv240806155D}
}

@ARTICLE{owen_SWA2020,
       author = {{Owen}, C.~J. and {Bruno}, R. and {Livi}, S. and {Louarn}, P. and {Al Janabi}, K. and {Allegrini}, F. and {Amoros}, C. and {Baruah}, R. and {Barthe}, A. and {Berthomier}, M. and {Bordon}, S. and {Brockley-Blatt}, C. and {Brysbaert}, C. and {Capuano}, G. and {Collier}, M. and {DeMarco}, R. and {Fedorov}, A. and {Ford}, J. and {Fortunato}, V. and {Fratter}, I. and {Galvin}, A.~B. and {Hancock}, B. and {Heirtzler}, D. and {Kataria}, D. and {Kistler}, L. and {Lepri}, S.~T. and {Lewis}, G. and {Loeffler}, C. and {Marty}, W. and {Mathon}, R. and {Mayall}, A. and {Mele}, G. and {Ogasawara}, K. and {Orlandi}, M. and {Pacros}, A. and {Penou}, E. and {Persyn}, S. and {Petiot}, M. and {Phillips}, M. and {P{\v{r}}ech}, L. and {Raines}, J.~M. and {Reden}, M. and {Rouillard}, A.~P. and {Rousseau}, A. and {Rubiella}, J. and {Seran}, H. and {Spencer}, A. and {Thomas}, J.~W. and {Trevino}, J. and {Verscharen}, D. and {Wurz}, P. and {Alapide}, A. and {Amoruso}, L. and {Andr{\'e}}, N. and {Anekallu}, C. and {Arciuli}, V. and {Arnett}, K.~L. and {Ascolese}, R. and {Bancroft}, C. and {Bland}, P. and {Brysch}, M. and {Calvanese}, R. and {Castronuovo}, M. and {{\v{C}}erm{\'a}k}, I. and {Chornay}, D. and {Clemens}, S. and {Coker}, J. and {Collinson}, G. and {D'Amicis}, R. and {Dandouras}, I. and {Darnley}, R. and {Davies}, D. and {Davison}, G. and {De Los Santos}, A. and {Devoto}, P. and {Dirks}, G. and {Edlund}, E. and {Fazakerley}, A. and {Ferris}, M. and {Frost}, C. and {Fruit}, G. and {Garat}, C. and {G{\'e}not}, V. and {Gibson}, W. and {Gilbert}, J.~A. and {de Giosa}, V. and {Gradone}, S. and {Hailey}, M. and {Horbury}, T.~S. and {Hunt}, T. and {Jacquey}, C. and {Johnson}, M. and {Lavraud}, B. and {Lawrenson}, A. and {Leblanc}, F. and {Lockhart}, W. and {Maksimovic}, M. and {Malpus}, A. and {Marcucci}, F. and {Mazelle}, C. and {Monti}, F. and {Myers}, S. and {Nguyen}, T. and {Rodriguez-Pacheco}, J. and {Phillips}, I. and {Popecki}, M. and {Rees}, K. and {Rogacki}, S.~A. and {Ruane}, K. and {Rust}, D. and {Salatti}, M. and {Sauvaud}, J.~A. and {Stakhiv}, M.~O. and {Stange}, J. and {Stubbs}, T. and {Taylor}, T. and {Techer}, J. -D. and {Terrier}, G. and {Thibodeaux}, R. and {Urdiales}, C. and {Varsani}, A. and {Walsh}, A.~P. and {Watson}, G. and {Wheeler}, P. and {Willis}, G. and {Wimmer-Schweingruber}, R.~F. and {Winter}, B. and {Yardley}, J. and {Zouganelis}, I.},
        title = "{The Solar Orbiter Solar Wind Analyser (SWA) suite}",
      journal = {\aap},
         year = 2020,
        month = oct,
       volume = {642},
          eid = {A16},
        pages = {A16},
          doi = {10.1051/0004-6361/201937259},
       adsurl = {https://ui.adsabs.harvard.edu/abs/2020A&A...642A..16O}
}

@article{parker1958,
       author = {{Parker}, E.~N.},
        title = "{Dynamics of the Interplanetary Gas and Magnetic Fields.}",
      journal = {\apj},
         year = 1958,
        month = nov,
       volume = {128},
        pages = {664},
          doi = {10.1086/146579},
       adsurl = {https://ui.adsabs.harvard.edu/abs/1958ApJ...128..664P}
}

@ARTICLE{weber_davis1967,
       author = {{Weber}, Edmund J. and {Davis}, Leverett, Jr.},
        title = "{The Angular Momentum of the Solar Wind}",
      journal = {\apj},
         year = 1967,
        month = apr,
       volume = {148},
        pages = {217-227},
          doi = {10.1086/149138},
       adsurl = {https://ui.adsabs.harvard.edu/abs/1967ApJ...148..217W}
}

@ARTICLE{reville2020,
       author = {{R{\'e}ville}, Victor and {Velli}, Marco and {Panasenco}, Olga and {Tenerani}, Anna and {Shi}, Chen and {Badman}, Samuel T. and {Bale}, Stuart D. and {Kasper}, J.~C. and {Stevens}, Michael L. and {Korreck}, Kelly E. and {Bonnell}, J.~W. and {Case}, Anthony W. and {de Wit}, Thierry Dudok and {Goetz}, Keith and {Harvey}, Peter R. and {Larson}, Davin E. and {Livi}, Roberto and {Malaspina}, David M. and {MacDowall}, Robert J. and {Pulupa}, Marc and {Whittlesey}, Phyllis L.},
        title = "{The Role of Alfv{\'e}n Wave Dynamics on the Large-scale Properties of the Solar Wind: Comparing an MHD Simulation with Parker Solar Probe E1 Data}",
      journal = {\apjs},
         year = 2020,
        month = feb,
       volume = {246},
       number = {2},
          eid = {24},
        pages = {24},
          doi = {10.3847/1538-4365/ab4fef},
archivePrefix = {arXiv},
       eprint = {1912.03777},
 primaryClass = {astro-ph.SR},
       adsurl = {https://ui.adsabs.harvard.edu/abs/2020ApJS..246...24R}
}

@ARTICLE{halekas2022,
       author = {{Halekas}, J.~S. and {Whittlesey}, P. and {Larson}, D.~E. and {Maksimovic}, M. and {Livi}, R. and {Berthomier}, M. and {Kasper}, J.~C. and {Case}, A.~W. and {Stevens}, M.~L. and {Bale}, S.~D. and {MacDowall}, R.~J. and {Pulupa}, M.~P.},
        title = "{The Radial Evolution of the Solar Wind as Organized by Electron Distribution Parameters}",
      journal = {\apj},
         year = 2022,
        month = sep,
       volume = {936},
       number = {1},
          eid = {53},
        pages = {53},
          doi = {10.3847/1538-4357/ac85b8},
archivePrefix = {arXiv},
       eprint = {2207.06563},
 primaryClass = {astro-ph.SR},
       adsurl = {https://ui.adsabs.harvard.edu/abs/2022ApJ...936...53H}
}

@ARTICLE{schatten1969,
       author = {{Schatten}, Kenneth H. and {Wilcox}, John M. and {Ness}, Norman F.},
        title = "{A model of interplanetary and coronal magnetic fields}",
      journal = {\solphys},
         year = 1969,
        month = mar,
       volume = {6},
       number = {3},
        pages = {442-455},
          doi = {10.1007/BF00146478},
       adsurl = {https://ui.adsabs.harvard.edu/abs/1969SoPh....6..442S}
}

@ARTICLE{rouillard2020,
       author = {{Rouillard}, Alexis P. and {Kouloumvakos}, Athanasios and {Vourlidas}, Angelos and {Kasper}, Justin and {Bale}, Stuart and {Raouafi}, Nour-Edine and {Lavraud}, Benoit and {Howard}, Russell A. and {Stenborg}, Guillermo and {Stevens}, Michael and {Poirier}, Nicolas and {Davies}, Jackie A. and {Hess}, Phillip and {Higginson}, Aleida K. and {Lavarra}, Michael and {Viall}, Nicholeen M. and {Korreck}, Kelly and {Pinto}, Rui F. and {Griton}, L{\'e}a and {R{\'e}ville}, Victor and {Louarn}, Philippe and {Wu}, Yihong and {Dalmasse}, K{\'e}vin and {G{\'e}not}, Vincent and {Case}, Anthony W. and {Whittlesey}, Phyllis and {Larson}, Davin and {Halekas}, Jasper S. and {Livi}, Roberto and {Goetz}, Keith and {Harvey}, Peter R. and {MacDowall}, Robert J. and {Malaspina}, D. and {Pulupa}, M. and {Bonnell}, J. and {de Witt}, T. Dudok and {Penou}, Emmanuel},
        title = "{Relating Streamer Flows to Density and Magnetic Structures at the Parker Solar Probe}",
      journal = {\apjs},
         year = 2020,
        month = feb,
       volume = {246},
       number = {2},
          eid = {37},
        pages = {37},
          doi = {10.3847/1538-4365/ab579a},
archivePrefix = {arXiv},
       eprint = {2001.01993},
 primaryClass = {astro-ph.SR},
       adsurl = {https://ui.adsabs.harvard.edu/abs/2020ApJS..246...37R}
}

@ARTICLE{Badman2020,
       author = {{Badman}, Samuel T. and {Bale}, Stuart D. and {Mart{\'\i}nez Oliveros}, Juan C. and {Panasenco}, Olga and {Velli}, Marco and {Stansby}, David and {Buitrago-Casas}, Juan C. and {R{\'e}ville}, Victor and {Bonnell}, John W. and {Case}, Anthony W. and {Dudok de Wit}, Thierry and {Goetz}, Keith and {Harvey}, Peter R. and {Kasper}, Justin C. and {Korreck}, Kelly E. and {Larson}, Davin E. and {Livi}, Roberto and {MacDowall}, Robert J. and {Malaspina}, David M. and {Pulupa}, Marc and {Stevens}, Michael L. and {Whittlesey}, Phyllis L.},
        title = "{Magnetic Connectivity of the Ecliptic Plane within 0.5 au: Potential Field Source Surface Modeling of the First Parker Solar Probe Encounter}",
      journal = {\apjs},
         year = 2020,
        month = feb,
       volume = {246},
       number = {2},
          eid = {23},
        pages = {23},
          doi = {10.3847/1538-4365/ab4da7},
archivePrefix = {arXiv},
       eprint = {1912.02244},
 primaryClass = {astro-ph.SR},
       adsurl = {https://ui.adsabs.harvard.edu/abs/2020ApJS..246...23B}
}

@ARTICLE{muller2020,
       author = {{M{\"u}ller}, D. and {St. Cyr}, O.~C. and {Zouganelis}, I. and {Gilbert}, H.~R. and {Marsden}, R. and {Nieves-Chinchilla}, T. and {Antonucci}, E. and {Auch{\`e}re}, F. and {Berghmans}, D. and {Horbury}, T.~S. and {Howard}, R.~A. and {Krucker}, S. and {Maksimovic}, M. and {Owen}, C.~J. and {Rochus}, P. and {Rodriguez-Pacheco}, J. and {Romoli}, M. and {Solanki}, S.~K. and {Bruno}, R. and {Carlsson}, M. and {Fludra}, A. and {Harra}, L. and {Hassler}, D.~M. and {Livi}, S. and {Louarn}, P. and {Peter}, H. and {Sch{\"u}hle}, U. and {Teriaca}, L. and {del Toro Iniesta}, J.~C. and {Wimmer-Schweingruber}, R.~F. and {Marsch}, E. and {Velli}, M. and {De Groof}, A. and {Walsh}, A. and {Williams}, D.},
        title = "{The Solar Orbiter mission. Science overview}",
      journal = {\aap},
         year = 2020,
        month = oct,
       volume = {642},
          eid = {A1},
        pages = {A1},
          doi = {10.1051/0004-6361/202038467},
archivePrefix = {arXiv},
       eprint = {2009.00861},
 primaryClass = {astro-ph.SR},
       adsurl = {https://ui.adsabs.harvard.edu/abs/2020A&A...642A...1M}
}

@ARTICLE{arden2014,
       author = {{Arden}, W.~M. and {Norton}, A.~A. and {Sun}, X.},
        title = "{A ``breathing'' source surface for cycles 23 and 24}",
      journal = {Journal of Geophysical Research (Space Physics)},
         year = 2014,
        month = mar,
       volume = {119},
       number = {3},
        pages = {1476-1485},
          doi = {10.1002/2013JA019464},
       adsurl = {https://ui.adsabs.harvard.edu/abs/2014JGRA..119.1476A}
}

@ARTICLE{belcher1971,
       author = {{Belcher}, J.~W.},
        title = "{ALFV{\'E}NIC Wave Pressures and the Solar Wind}",
      journal = {\apj},
         year = 1971,
        month = sep,
       volume = {168},
        pages = {509},
          doi = {10.1086/151105},
       adsurl = {https://ui.adsabs.harvard.edu/abs/1971ApJ...168..509B}
}

@ARTICLE{verdini2007,
       author = {{Verdini}, Andrea and {Velli}, Marco},
        title = "{Alfv{\'e}n Waves and Turbulence in the Solar Atmosphere and Solar Wind}",
      journal = {\apj},
         year = 2007,
        month = jun,
       volume = {662},
       number = {1},
        pages = {669-676},
          doi = {10.1086/510710},
archivePrefix = {arXiv},
       eprint = {astro-ph/0702205},
 primaryClass = {astro-ph},
       adsurl = {https://ui.adsabs.harvard.edu/abs/2007ApJ...662..669V}
}

@ARTICLE{chandran2009,
       author = {{Chandran}, Benjamin D.~G. and {Hollweg}, Joseph V.},
        title = "{Alfv{\'e}n Wave Reflection and Turbulent Heating in the Solar Wind from 1 Solar Radius to 1 AU: An Analytical Treatment}",
      journal = {\apj},
         year = 2009,
        month = dec,
       volume = {707},
       number = {2},
        pages = {1659-1667},
          doi = {10.1088/0004-637X/707/2/1659},
archivePrefix = {arXiv},
       eprint = {0911.1068},
 primaryClass = {astro-ph.SR},
       adsurl = {https://ui.adsabs.harvard.edu/abs/2009ApJ...707.1659C}
}

@INPROCEEDINGS{adapt_ref2013,
       author = {{Arge}, C. Nick and {Henney}, Carl J. and {Hernandez}, Irene Gonzalez and {Toussaint}, W. Alex and {Koller}, Josef and {Godinez}, Humberto C.},
        title = "{Modeling the corona and solar wind using ADAPT maps that include far-side observations}",
    booktitle = {Solar Wind 13},
         year = 2013,
       editor = {{Zank}, Gary P. and {Borovsky}, Joe and {Bruno}, Roberto and {Cirtain}, Jonathan and {Cranmer}, Steve and {Elliott}, Heather and {Giacalone}, Joe and {Gonzalez}, Walter and {Li}, Gang and {Marsch}, Eckart and {Moebius}, Ebehard and {Pogorelov}, Nick and {Spann}, Jim and {Verkhoglyadova}, Olga},
       series = {American Institute of Physics Conference Series},
       volume = {1539},
        month = jun,
        pages = {11-14},
          doi = {10.1063/1.4810977},
       adsurl = {https://ui.adsabs.harvard.edu/abs/2013AIPC.1539...11A}
}

@ARTICLE{poduval2004,
       author = {{Poduval}, Bala and {Zhao}, Xue Pu},
        title = "{Discrepancies in the prediction of solar wind using potential field source surface model: An investigation of possible sources}",
      journal = {Journal of Geophysical Research (Space Physics)},
         year = 2004,
        month = aug,
       volume = {109},
       number = {A8},
          eid = {A08102},
        pages = {A08102},
          doi = {10.1029/2004JA010384},
       adsurl = {https://ui.adsabs.harvard.edu/abs/2004JGRA..109.8102P}
}

@ARTICLE{Li2021,
       author = {{Li}, Huichao and {Feng}, Xueshang and {Wei}, Fengsi},
        title = "{Comparison of Synoptic Maps and PFSS Solutions for The Declining Phase of Solar Cycle 24}",
      journal = {Journal of Geophysical Research (Space Physics)},
         year = 2021,
        month = mar,
       volume = {126},
       number = {3},
          eid = {e28870},
        pages = {e28870},
          doi = {10.1029/2020JA028870},
       adsurl = {https://ui.adsabs.harvard.edu/abs/2021JGRA..12628870L}
}

@ARTICLE{Horbury_MAG2020,
       author = {{Horbury}, T.~S. and {O'Brien}, H. and {Carrasco Blazquez}, I. and {Bendyk}, M. and {Brown}, P. and {Hudson}, R. and {Evans}, V. and {Oddy}, T.~M. and {Carr}, C.~M. and {Beek}, T.~J. and {Cupido}, E. and {Bhattacharya}, S. and {Dominguez}, J. -A. and {Matthews}, L. and {Myklebust}, V.~R. and {Whiteside}, B. and {Bale}, S.~D. and {Baumjohann}, W. and {Burgess}, D. and {Carbone}, V. and {Cargill}, P. and {Eastwood}, J. and {Erd{\"o}s}, G. and {Fletcher}, L. and {Forsyth}, R. and {Giacalone}, J. and {Glassmeier}, K. -H. and {Goldstein}, M.~L. and {Hoeksema}, T. and {Lockwood}, M. and {Magnes}, W. and {Maksimovic}, M. and {Marsch}, E. and {Matthaeus}, W.~H. and {Murphy}, N. and {Nakariakov}, V.~M. and {Owen}, C.~J. and {Owens}, M. and {Rodriguez-Pacheco}, J. and {Richter}, I. and {Riley}, P. and {Russell}, C.~T. and {Schwartz}, S. and {Vainio}, R. and {Velli}, M. and {Vennerstrom}, S. and {Walsh}, R. and {Wimmer-Schweingruber}, R.~F. and {Zank}, G. and {M{\"u}ller}, D. and {Zouganelis}, I. and {Walsh}, A.~P.},
        title = "{The Solar Orbiter magnetometer}",
      journal = {\aap},
         year = 2020,
        month = oct,
       volume = {642},
          eid = {A9},
        pages = {A9},
          doi = {10.1051/0004-6361/201937257},
       adsurl = {https://ui.adsabs.harvard.edu/abs/2020A&A...642A...9H}
}

@ARTICLE{Alazraki1971,
       author = {{Alazraki}, G. and {Couturier}, P.},
        title = "{Solar Wind Accejeration Caused by the Gradient of Alfven Wave Pressure}",
      journal = {\aap},
         year = 1971,
        month = aug,
       volume = {13},
        pages = {380},
       adsurl = {https://ui.adsabs.harvard.edu/abs/1971A&A....13..380A}
}

@ARTICLE{Rivera2024,
       author = {{Rivera}, Yeimy J. and {Badman}, Samuel T. and {Stevens}, Michael L. and {Verniero}, Jaye L. and {Stawarz}, Julia E. and {Shi}, Chen and {Raines}, Jim M. and {Paulson}, Kristoff W. and {Owen}, Christopher J. and {Niembro}, Tatiana and {Louarn}, Philippe and {Livi}, Stefano A. and {Lepri}, Susan T. and {Kasper}, Justin C. and {Horbury}, Timothy S. and {Halekas}, Jasper S. and {Dewey}, Ryan M. and {De Marco}, Rossana and {Bale}, Stuart D.},
        title = "{In situ observations of large-amplitude Alfv{\'e}n waves heating and accelerating the solar wind}",
      journal = {Science},
         year = 2024,
        month = aug,
       volume = {385},
       number = {6712},
        pages = {962-966},
          doi = {10.1126/science.adk6953},
archivePrefix = {arXiv},
       eprint = {2409.00267},
 primaryClass = {astro-ph.SR},
       adsurl = {https://ui.adsabs.harvard.edu/abs/2024Sci...385..962R}
}

@ARTICLE{Rivera2025,
       author = {{Rivera}, Yeimy J. and {Badman}, Samuel T. and {Verniero}, J.~L. and {Varesano}, Tania and {Stevens}, Michael L. and {Stawarz}, Julia E. and {Reeves}, Katharine K. and {Raines}, Jim M. and {Raymond}, John C. and {Owen}, Christopher J. and {Livi}, Stefano A. and {Lepri}, Susan T. and {Landi}, Enrico and {Halekas}, Jasper. S. and {Ervin}, Tamar and {Dewey}, Ryan M. and {De Marco}, Rossana and {D'Amicis}, Raffaella and {Dakeyo}, Jean-Baptiste and {Bale}, Stuart D. and {Alterman}, B.~L.},
        title = "{Differentiating the Acceleration Mechanisms in the Slow and Alfv{\'e}nic Slow Solar Wind}",
      journal = {\apj},
         year = 2025,
        month = feb,
       volume = {980},
       number = {1},
          eid = {70},
        pages = {70},
          doi = {10.3847/1538-4357/ada699},
archivePrefix = {arXiv},
       eprint = {2501.02163},
 primaryClass = {astro-ph.SR},
       adsurl = {https://ui.adsabs.harvard.edu/abs/2025ApJ...980...70R}
}

@ARTICLE{Telloni2021,
       author = {{Telloni}, Daniele and {Sorriso-Valvo}, Luca and {Woodham}, Lloyd D. and {Panasenco}, Olga and {Velli}, Marco and {Carbone}, Francesco and {Zank}, Gary P. and {Bruno}, Roberto and {Perrone}, Denise and {Nakanotani}, Masaru and {Shi}, Chen and {D'Amicis}, Raffaella and {De Marco}, Rossana and {Jagarlamudi}, Vamsee K. and {Steinvall}, Konrad and {Marino}, Raffaele and {Adhikari}, Laxman and {Zhao}, Lingling and {Liang}, Haoming and {Tenerani}, Anna and {Laker}, Ronan and {Horbury}, Timothy S. and {Bale}, Stuart D. and {Pulupa}, Marc and {Malaspina}, David M. and {MacDowall}, Robert J. and {Goetz}, Keith and {de Wit}, Thierry Dudok and {Harvey}, Peter R. and {Kasper}, Justin C. and {Korreck}, Kelly E. and {Larson}, Davin and {Case}, Anthony W. and {Stevens}, Michael L. and {Whittlesey}, Phyllis and {Livi}, Roberto and {Owen}, Christopher J. and {Livi}, Stefano and {Louarn}, Philippe and {Antonucci}, Ester and {Romoli}, Marco and {O'Brien}, Helen and {Evans}, Vincent and {Angelini}, Virginia},
        title = "{Evolution of Solar Wind Turbulence from 0.1 to 1 au during the First Parker Solar Probe-Solar Orbiter Radial Alignment}",
      journal = {\apjl},
         year = 2021,
        month = may,
       volume = {912},
       number = {2},
          eid = {L21},
        pages = {L21},
          doi = {10.3847/2041-8213/abf7d1},
       adsurl = {https://ui.adsabs.harvard.edu/abs/2021ApJ...912L..21T}
}

@ARTICLE{Berriot2024,
       author = {{Berriot}, Etienne and {D{\'e}moulin}, Pascal and {Alexandrova}, Olga and {Zaslavsky}, Arnaud and {Maksimovic}, Milan},
        title = "{Identification of a single plasma parcel during a radial alignment of the Parker Solar Probe and Solar Orbiter}",
      journal = {\aap},
         year = 2024,
        month = jun,
       volume = {686},
          eid = {A114},
        pages = {A114},
          doi = {10.1051/0004-6361/202449285},
archivePrefix = {arXiv},
       eprint = {2402.12382},
 primaryClass = {astro-ph.SR},
       adsurl = {https://ui.adsabs.harvard.edu/abs/2024A&A...686A.114B}
}

@ARTICLE{Berriot2025,
       author = {{Berriot}, Etienne and {D{\'e}moulin}, Pascal and {Alexandrova}, Olga and {Zaslavsky}, Arnaud and {Maksimovic}, Milan and {Nicolaou}, Georgios},
        title = "{Radial Evolution of a Density Structure within a Solar Wind Magnetic Sector Boundary}",
      journal = {\apj},
         year = 2025,
        month = mar,
       volume = {981},
       number = {2},
          eid = {140},
        pages = {140},
          doi = {10.3847/1538-4357/adb39a},
archivePrefix = {arXiv},
       eprint = {2412.09395},
 primaryClass = {astro-ph.SR},
       adsurl = {https://ui.adsabs.harvard.edu/abs/2025ApJ...981..140B}
}

@ARTICLE{Ervin2024a,
       author = {{Ervin}, Tamar and {Bale}, Stuart D. and {Badman}, Samuel T. and {Rivera}, Yeimy J. and {Romeo}, Orlando and {Huang}, Jia and {Riley}, Pete and {Bowen}, Trevor A. and {Lepri}, Susan T. and {Dewey}, Ryan M.},
        title = "{Compositional Metrics of Fast and Slow Alfv{\'e}nic Solar Wind Emerging from Coronal Holes and Their Boundaries}",
      journal = {\apj},
         year = 2024,
        month = jul,
       volume = {969},
       number = {2},
          eid = {83},
        pages = {83},
          doi = {10.3847/1538-4357/ad4604},
archivePrefix = {arXiv},
       eprint = {2309.07949},
 primaryClass = {astro-ph.SR},
       adsurl = {https://ui.adsabs.harvard.edu/abs/2024ApJ...969...83E}
}

@ARTICLE{Ervin2024b,
       author = {{Ervin}, Tamar and {Bale}, Stuart D. and {Badman}, Samuel T. and {Bowen}, Trevor A. and {Riley}, Pete and {Paulson}, Kristoff and {Rivera}, Yeimy J. and {Romeo}, Orlando and {Sioulas}, Nikos and {Larson}, Davin and {Verniero}, Jaye L. and {Dewey}, Ryan M. and {Huang}, Jia},
        title = "{Near Subsonic Solar Wind Outflow from an Active Region}",
      journal = {\apj},
         year = 2024,
        month = sep,
       volume = {972},
       number = {1},
          eid = {129},
        pages = {129},
          doi = {10.3847/1538-4357/ad57c4},
archivePrefix = {arXiv},
       eprint = {2405.15844},
 primaryClass = {astro-ph.SR},
       adsurl = {https://ui.adsabs.harvard.edu/abs/2024ApJ...972..129E}
}

@ARTICLE{Alexandrova2013,
       author = {{Alexandrova}, O. and {Chen}, C.~H.~K. and {Sorriso-Valvo}, L. and {Horbury}, T.~S. and {Bale}, S.~D.},
        title = "{Solar Wind Turbulence and the Role of Ion Instabilities}",
      journal = {\ssr},
         year = 2013,
        month = oct,
       volume = {178},
       number = {2-4},
        pages = {101-139},
          doi = {10.1007/s11214-013-0004-8},
archivePrefix = {arXiv},
       eprint = {1306.5336},
 primaryClass = {astro-ph.SR},
       adsurl = {https://ui.adsabs.harvard.edu/abs/2013SSRv..178..101A}
}

@ARTICLE{Sioulas2023,
       author = {{Sioulas}, Nikos and {Huang}, Zesen and {Shi}, Chen and {Velli}, Marco and {Tenerani}, Anna and {Bowen}, Trevor A. and {Bale}, Stuart D. and {Huang}, Jia and {Vlahos}, Loukas and {Woodham}, L.~D. and {Horbury}, T.~S. and {de Wit}, Thierry Dudok and {Larson}, Davin and {Kasper}, Justin and {Owen}, Christopher J. and {Stevens}, Michael L. and {Case}, Anthony and {Pulupa}, Marc and {Malaspina}, David M. and {Bonnell}, J.~W. and {Livi}, Roberto and {Goetz}, Keith and {Harvey}, Peter R. and {MacDowall}, Robert J. and {Maksimovi{\'c}}, Milan and {Louarn}, P. and {Fedorov}, A.},
        title = "{Magnetic Field Spectral Evolution in the Inner Heliosphere}",
      journal = {\apjl},
         year = 2023,
        month = jan,
       volume = {943},
       number = {1},
          eid = {L8},
        pages = {L8},
          doi = {10.3847/2041-8213/acaeff},
archivePrefix = {arXiv},
       eprint = {2209.02451},
 primaryClass = {astro-ph.SR},
       adsurl = {https://ui.adsabs.harvard.edu/abs/2023ApJ...943L...8S}
}

@ARTICLE{Huang2025,
       author = {{Huang}, Zesen and {Velli}, Marco and {Chandran}, B.~D.~G. and {Shi}, Chen and {Ding}, Yuliang and {Matteini}, Lorenzo and {Choi}, Kyung-Eun},
        title = "{Two Types of $1/f$ Range in Solar Wind Turbulence}",
      journal = {arXiv e-prints},
         year = 2025,
        month = jun,
          eid = {arXiv:2506.17523},
        pages = {arXiv:2506.17523},
          doi = {10.48550/arXiv.2506.17523},
archivePrefix = {arXiv},
       eprint = {2506.17523},
 primaryClass = {astro-ph.SR},
       adsurl = {https://ui.adsabs.harvard.edu/abs/2025arXiv250617523H}
}

@ARTICLE{QTN_ref_2020,
       author = {{Moncuquet}, Michel and {Meyer-Vernet}, Nicole and {Issautier}, Karine and {Pulupa}, Marc and {Bonnell}, J.~W. and {Bale}, Stuart D. and {Dudok de Wit}, Thierry and {Goetz}, Keith and {Griton}, L{\'e}a and {Harvey}, Peter R. and {MacDowall}, Robert J. and {Maksimovic}, Milan and {Malaspina}, David M.},
        title = "{First In Situ Measurements of Electron Density and Temperature from Quasi-thermal Noise Spectroscopy with Parker Solar Probe/FIELDS}",
      journal = {\apjs},
         year = 2020,
        month = feb,
       volume = {246},
       number = {2},
          eid = {44},
        pages = {44},
          doi = {10.3847/1538-4365/ab5a84},
archivePrefix = {arXiv},
       eprint = {1912.02518},
 primaryClass = {astro-ph.SR},
       adsurl = {https://ui.adsabs.harvard.edu/abs/2020ApJS..246...44M}
}

@article{Moestl2020_icme_catalog,
author = "Christian Moestl and Emma Davies and Eva Weiler",
title = "{HELIO4CAST Interplanetary Coronal Mass Ejection Catalog v2.3}",
year = "2020",
month = "6",
url = "https://figshare.com/articles/dataset/HELCATS_Interplanetary_Coronal_Mass_Ejection_Catalog_v2_0/6356420",
doi = "10.6084/m9.figshare.6356420.v23"
}

@ARTICLE{Halekas2023,
       author = {{Halekas}, J.~S. and {Bale}, S.~D. and {Berthomier}, M. and {Chandran}, B.~D.~G. and {Drake}, J.~F. and {Kasper}, J.~C. and {Klein}, K.~G. and {Larson}, D.~E. and {Livi}, R. and {Pulupa}, M.~P. and {Stevens}, M.~L. and {Verniero}, J.~L. and {Whittlesey}, P.},
        title = "{Quantifying the Energy Budget in the Solar Wind from 13.3 to 100 Solar Radii}",
      journal = {\apj},
         year = 2023,
        month = jul,
       volume = {952},
       number = {1},
          eid = {26},
        pages = {26},
          doi = {10.3847/1538-4357/acd769},
archivePrefix = {arXiv},
       eprint = {2305.13424},
 primaryClass = {astro-ph.SR},
       adsurl = {https://ui.adsabs.harvard.edu/abs/2023ApJ...952...26H}
}

@ARTICLE{SPAN_E_ref_2020,
       author = {{Whittlesey}, Phyllis L. and {Larson}, Davin E. and {Kasper}, Justin C. and {Halekas}, Jasper and {Abatcha}, Mamuda and {Abiad}, Robert and {Berthomier}, M. and {Case}, A.~W. and {Chen}, Jianxin and {Curtis}, David W. and {Dalton}, Gregory and {Klein}, Kristopher G. and {Korreck}, Kelly E. and {Livi}, Roberto and {Ludlam}, Michael and {Marckwordt}, Mario and {Rahmati}, Ali and {Robinson}, Miles and {Slagle}, Amanda and {Stevens}, M.~L. and {Tiu}, Chris and {Verniero}, J.~L.},
        title = "{The Solar Probe ANalyzers{\textemdash}Electrons on the Parker Solar Probe}",
      journal = {\apjs},
         year = 2020,
        month = feb,
       volume = {246},
       number = {2},
          eid = {74},
        pages = {74},
          doi = {10.3847/1538-4365/ab7370},
archivePrefix = {arXiv},
       eprint = {2002.04080},
 primaryClass = {astro-ph.IM},
       adsurl = {https://ui.adsabs.harvard.edu/abs/2020ApJS..246...74W}
}

@ARTICLE{Silwal2025,
       author = {{Silwal}, Ashok and {Zhao}, Lingling and {Zhu}, Xingyu and {Sorriso-Valvo}, Luca and {Hadid}, Lina Z. and {Zank}, Gary P. and {Li}, Hui and {Badman}, Samuel T. and {Rivera}, Yeimy J. and {Gautam}, Sujan Prasad and {Karki}, Monika and {Alonso Guzman}, Juan G. and {M. Subashchandar}, Nibuna S. and {Jin}, Zeping},
        title = "{Evolution of Solar Wind Turbulence during Radial Alignment of Parker Solar Probe and Solar Orbiter in 2022 December}",
      journal = {\apjs},
         year = 2025,
        month = jun,
       volume = {278},
       number = {2},
          eid = {40},
        pages = {40},
          doi = {10.3847/1538-4365/add011},
       adsurl = {https://ui.adsabs.harvard.edu/abs/2025ApJS..278...40S}
}

@ARTICLE{Matteini2007,
       author = {{Matteini}, Lorenzo and {Landi}, Simone and {Hellinger}, Petr and {Pantellini}, Filippo and {Maksimovic}, Milan and {Velli}, Marco and {Goldstein}, Bruce E. and {Marsch}, Eckart},
        title = "{Evolution of the solar wind proton temperature anisotropy from 0.3 to 2.5 AU}",
      journal = {\grl},
         year = 2007,
        month = oct,
       volume = {34},
       number = {20},
          eid = {L20105},
        pages = {L20105},
          doi = {10.1029/2007GL030920},
       adsurl = {https://ui.adsabs.harvard.edu/abs/2007GeoRL..3420105M}
}

@ARTICLE{2021LiuAA,
       author = {{Liu}, M. and {Issautier}, K. and {Meyer-Vernet}, N. and {Moncuquet}, M. and {Maksimovic}, M. and {Halekas}, J.~S. and {Huang}, J. and {Griton}, L. and {Bale}, S. and {Bonnell}, J.~W. and {Case}, A.~W. and {Goetz}, K. and {Harvey}, P.~R. and {Kasper}, J.~C. and {MacDowall}, R.~J. and {Malaspina}, D.~M. and {Pulupa}, M. and {Stevens}, M.~L.},
        title = "{Solar wind energy flux observations in the inner heliosphere: first results from Parker Solar Probe}",
      journal = {\aap},
         year = 2021,
        month = jun,
       volume = {650},
          eid = {A14},
        pages = {A14},
          doi = {10.1051/0004-6361/202039615},
archivePrefix = {arXiv},
       eprint = {2101.03121},
 primaryClass = {astro-ph.SR},
       adsurl = {https://ui.adsabs.harvard.edu/abs/2021A&A...650A..14L}
}

@ARTICLE{2023LiuAA,
       author = {{Liu}, M. and {Issautier}, K. and {Moncuquet}, M. and {Meyer-Vernet}, N. and {Maksimovic}, M. and {Huang}, J. and {Martinovic}, M.~M. and {Griton}, L. and {Chrysaphi}, N. and {Jagarlamudi}, V.~K. and {Bale}, S.~D. and {Pulupa}, M. and {Kasper}, J.~C. and {Stevens}, M.~L.},
        title = "{Total electron temperature derived from quasi-thermal noise spectroscopy in the pristine solar wind from Parker Solar Probe observations}",
      journal = {\aap},
         year = 2023,
        month = jun,
       volume = {674},
          eid = {A49},
        pages = {A49},
          doi = {10.1051/0004-6361/202245450},
archivePrefix = {arXiv},
       eprint = {2303.11035},
 primaryClass = {astro-ph.SR},
       adsurl = {https://ui.adsabs.harvard.edu/abs/2023A&A...674A..49L}
}

@INPROCEEDINGS{Arge2010,
       author = {{Arge}, C. Nick and {Henney}, Carl J. and {Koller}, Josef and {Compeau}, C. Rich and {Young}, Shawn and {MacKenzie}, David and {Fay}, Alex and {Harvey}, John W.},
        title = "{Air Force Data Assimilative Photospheric Flux Transport (ADAPT) Model}",
    booktitle = {Twelfth International Solar Wind Conference},
         year = 2010,
       editor = {{Maksimovic}, M. and {Issautier}, K. and {Meyer-Vernet}, N. and {Moncuquet}, M. and {Pantellini}, F.},
       series = {American Institute of Physics Conference Series},
       volume = {1216},
        month = mar,
    publisher = {AIP},
        pages = {343-346},
          doi = {10.1063/1.3395870},
       adsurl = {https://ui.adsabs.harvard.edu/abs/2010AIPC.1216..343A}
}

@ARTICLE{Telloni2023,
       author = {{Telloni}, Daniele},
        title = "{Spacecraft radial alignments for investigations of the evolution of solar wind turbulence: A review}",
      journal = {Journal of Atmospheric and Solar-Terrestrial Physics},
         year = 2023,
        month = jan,
       volume = {242},
          eid = {105999},
        pages = {105999},
          doi = {10.1016/j.jastp.2022.105999},
       adsurl = {https://ui.adsabs.harvard.edu/abs/2023JASTP.24205999T}
}

@ARTICLE{Sioulas2025b,
       author = {{Sioulas}, Nikos and {Velli}, Marco and {Shi}, Chen and {Bowen}, Trevor A. and {Mallet}, Alfred and {Verdini}, Andrea and {Chandran}, B.~D.~G. and {Tenerani}, Anna and {Dakeyo}, Jean-Baptiste and {Bale}, Stuart D. and {Larson}, Davin and {Halekas}, Jasper S. and {Matteini}, Lorenzo and {R{\'e}ville}, Victor and {Chen}, C.~H.~K. and {Romeo}, Orlando M. and {Liu}, Mingzhe and {Livi}, Roberto and {Rahmati}, Ali and {Whittlesey}, P.~L.},
        title = "{On the Propagation and Damping of Alfvenic Fluctuations in the Outer Solar Corona and Solar Wind}",
      journal = {arXiv e-prints},
         year = 2025,
        month = oct,
          eid = {arXiv:2510.10106},
        pages = {arXiv:2510.10106},
          doi = {10.48550/arXiv.2510.10106},
archivePrefix = {arXiv},
       eprint = {2510.10106},
 primaryClass = {astro-ph.SR},
       adsurl = {https://ui.adsabs.harvard.edu/abs/2025arXiv251010106S}
}

@ARTICLE{Damicis2021,
       author = {{D'Amicis}, R. and {Alielden}, K. and {Perrone}, D. and {Bruno}, R. and {Telloni}, D. and {Raines}, J.~M. and {Lepri}, S.~T. and {Zhao}, L.},
        title = "{Solar wind Alfv{\'e}nicity during solar cycle 23 and 24. Perspective for future observations with Parker Solar Probe and Solar Orbiter}",
      journal = {\aap},
         year = 2021,
        month = oct,
       volume = {654},
          eid = {A111},
        pages = {A111},
          doi = {10.1051/0004-6361/202140600},
       adsurl = {https://ui.adsabs.harvard.edu/abs/2021A&A...654A.111D}
}

@ARTICLE{Livi2022,
       author = {{Livi}, Roberto and {Larson}, Davin E. and {Kasper}, Justin C. and {Abiad}, Robert and {Case}, A.~W. and {Klein}, Kristopher G. and {Curtis}, David W. and {Dalton}, Gregory and {Stevens}, Michael and {Korreck}, Kelly E. and {Ho}, George and {Robinson}, Miles and {Tiu}, Chris and {Whittlesey}, Phyllis L. and {Verniero}, Jaye L. and {Halekas}, Jasper and {McFadden}, James and {Marckwordt}, Mario and {Slagle}, Amanda and {Abatcha}, Mamuda and {Rahmati}, Ali and {McManus}, Michael D.},
        title = "{The Solar Probe ANalyzer-Ions on the Parker Solar Probe}",
      journal = {\apj},
         year = 2022,
        month = oct,
       volume = {938},
       number = {2},
          eid = {138},
        pages = {138},
          doi = {10.3847/1538-4357/ac93f5},
       adsurl = {https://ui.adsabs.harvard.edu/abs/2022ApJ...938..138L}
}
\bibliographystyle{aasjournal}

\appendix

\section{Data}
\label{appendix:sec_data}

\subsection{Observations Pre-processing}
\label{appendix:sec_data_preprocessing}
\rev{
The field of view (FOV) of the SPAN-I instrument is affected by the thermal shield, as well as by the instrument orientation and the spacecraft velocity relative to the solar wind bulk flow. As a result, the velocity distribution function (VDF) is not always fully contained within the instrument FOV. Time intervals for which the VDF is significantly truncated are therefore removed. 
\\
This occurs when either the proton density $n_p$ decreases by more than 10\% relative to its daily average value, or when the maximum of the VDF energy flux is located above $170^\circ$ in the $\phi$ coordinate (based on the \textsc{EFLUX\_VS\_PHI} variable).\\
The total proton temperature measured by SPAN-I is adjusted to better match the radial temperature measured by the Solar Probe Cup (SPC), which has a more favorable FOV in the radial direction. A previous comparison between SPAN-I and SPC measurements showed that the radial temperature component satisfies $T_{zz} = T_{r|\spani} \approx 2 \: T_{r|\spc}$ \citep{dakeyo2022}. \\
We therefore correct the SPAN-I radial temperature by applying a factor of $1/2$ to ensure consistency with SPC measurements. In addition, the $T_{yy}$ component of the temperature tensor is subject to larger uncertainties (private communication with the SPAN-I team). Assuming that the solar wind is approximately gyrotropic \citep{Matteini2007}, we take $T_{xx} = T_{yy}$, which leads to the following expression for the total proton temperature:
$T_{\spani} = (2T_{xx} + T_{zz}/2)/3$.\\
The instrumental uncertainties for PSP data are approximately 3\% for the bulk speed and 10--15\% for the total proton temperature from SPAN-I; 10\% for the density and $\sim$20\% for the electron temperature derived from QTN observations of FIELDS \citep{QTN_ref_2020,2023LiuAA,2021LiuAA}; and 10--15\% for the electron temperature from SPAN-E (private communication with the SPAN-E team).
Instrumental uncertainties for SO are not explicitly provided. However, given that the FOV of PAS is more favorable than that of SPAN-I, and assuming similar performance for the MAG and FIELDS instruments, we adopt PSP instrumental uncertainties as a proxy for SO observations.\\
Interplanetary coronal mass ejections (ICMEs) are removed using the ICMECAT catalog \citep{Moestl2020_icme_catalog} for both PSP and SO datasets, and by applying the ICME identification criteria of \cite{Elliott2012temporal} to SO data, following the approach of \cite{dakeyo2024b}. Data are excluded when at least one of the following conditions on the plasma $\beta$ and proton temperature $\Tp$ is satisfied:
$(i)$ $\beta < 0.1$,  
$(ii)$ $\Tp / T_{ex} < 0.5$,  
where $T_{ex}$ is the expected temperature given by the scaling law  
$T_{ex}~=~486.5 \times u - 1.2476\times 10^5K$. In this formulation, $\Tp$ is rescaled with heliocentric distance using the predicted solar wind temperature $T_{ex}$.\\
ICMEs are assumed to have a minimum duration of 6 hours. In addition, we remove data within 24~hours prior to and 15~hours following each detected ICME. Finally, wind intervals with speeds exceeding 800~km/s are also excluded, as they are considered likely ICME-related.\\
The high-cadence datasets from PSP and SO are preprocessed using the same methodology. In the context of the backmapping algorithm, 30-minute averages are used solely to estimate the source regions of the \textit{in situ} solar wind observations. For the analysis of wind properties, plasma parameters are averaged over an adaptive time interval that depends on the radial distance $r$. This interval is chosen to match the large-scale fluctuation regime near the injection frequency $\finj$, corresponding to the low-frequency end of the inertial range.\\
We adopt the radial scaling $\finj \propto r^{-1}$ \citep{Sioulas2023, Huang2025}, and define:
\begin{align}
    \finj = 10^4 \times  (r / r_{1au})^{-1}.
\end{align}
At 1~au, this yields $\finj = 10^{-4}$~Hz, consistent with observations \citep{Alexandrova2013}. The adaptive time interval $\Tinj = 1/\finj$ is then used to compute averaged plasma properties.\\
These quantities are subsequently synchronized with the 30-minute mapping cadence, associating each stream source with \textit{in situ} measurements averaged over the injection scale.
}
\subsection{Fluctuations Computation}
\label{appendix:subsec_fluctuation_comput}
\rev{
The $J$-component fluctuation $\delta X_J$ of a given field $\mathbf{X}$ (either $\mathbf{v}$ or $\mathbf{B}$) is computed as the standard deviation of the fluctuations over an adaptive time interval of duration $\Tinj$:
\begin{align}
    \delta X_{J} = \: \bigg< \: \sqrt{ \frac{ \sum_{i=1}^{n} \: (X_{J}^{(i)} \: - <X_{J}>_{\Tinj})^2    }{n} } \: \bigg>_{\Tinj}
    \qquad \text{and} \qquad
    \delta X ~=~ \sqrt{\delta X_R^2 +\delta X_T^2 + \delta X_N^2 }, \label{eq_express_comput_dX}
\end{align}
where $n$ is the number of measurements within $\Tinj$, and $J$ denotes each of the $(R,T,N)$ components.
The total fluctuation $\delta X$ is defined as the amplitude of the vector fluctuation $\delta \mathbf{X}$.  
}

\section{Magnetic connectivity and backmapping specification}
\label{appendix:sec_backmapping}

\subsection{Backmapping technique}
\label{appendix:subsec_backmapping_technique}

The source alignment require to operate the magnetic connectivity of the observations to the photosphere. 
To connect our in-situ observations to their estimated solar atmospheric origin, we combine a magnetic field lines tracing (from the probe location) along the Parker spiral \citep{parker1958}, with a potential magnetic field reconstruction of the solar atmosphere \citep{schatten1969}. Since the magnetic field lines and the plasma streamline (i.e. plasma trajectory in the solar rotating frame) are aligned due to the ``frozen-in" plasma condition and stationary hypothesis, tracing back a stream magnetic origins allows to estimate its source at the Sun \citep{rouillard2020, Badman2020}. The backmapping process is a two step method. We follow the exact same technique as in \cite{dakeyo2024b}.
First, the solar wind trajectory (from the probe) is radially back-modeled considering estimation of the wind acceleration and corotation \citep{weber_davis1967, dakeyo2024a} based on isopoly models, that are \rev{Parker-like} models assuming an isothermal evolution of the solar wind near the Sun below the radial distance $r_{iso}$ (ranging from $\sim 3\, \rs$ and $\sim 20\, \rs$ depending the wind speed), and a polytropic evolution at larger distances.
Then getting close enough to the Sun below the so called "source surface" fixed at $\rss = 2.5\,\rs$ in our case (where the \rev{Parker} spiral modeling is not valid anymore) \citep{schatten1969}), 
we compute an extrapolation of the photospheric field using a potential field source surface (PFSS) model, between the source surface ($\rss$) and the photosphere ($\rs$). Taking the PFSS reconstructed field line rooted at the photosphere that connects to the in-situ spiral at $\rss$, gives an estimation of the solar wind source for a given in-situ measurement. \\
The mapping time interval is taken every 30 minutes according to averaged plasma properties (velocity, density, magnetic field). Each mapping is associated to an isopoly model that represent the plasma acceleration and corotation. Then, we compute the radial and tangential velocity profiles \citep{weber_davis1967,dakeyo2024a}.
The isopoly modeling follows \rev{Parker} equation \citep{parker1958}, with two distinct thermal regimes. It estimates an average constant temperature in the near Sun region, then a polytropic decreasing temperature in the interplanetary medium fitted to the actual observed decrease. This is done for several solar wind population. We take as reference the five wind population speed profiles, derived by \cite{dakeyo2022}. They present
1 au velocity of 292, 354, 406, 488 and 634~km/s. We interpolate (or extrapolate if the target speed is lower or faster than \rev{that of the reference population}) an isopoly profile to match the speed observed at a given probe location. 

The PFSS technique is operated at the same source surface height ($\rss = 2.5\,\rs$) for all the mappings. We select $\rss$ as the typical observed average height for magnetic field opening. Our PFSS algorithm uses a spherical harmonic decomposition of the ADAPT magnetogram following \cite{schatten1969}. The order of spherical harmonics $l$ is set to $l=20$. This constitutes an intermediate resolution compared with the resolution of the magnetogram, allowing to capture the complexity of the corona on the small and large scales 
while limiting artificial artifacts due to the decomposition itself \citep{ dakeyo2024b}.
\\

The streamline and PFSS modeling techniques are applied towards the Sun from PSP and SO spacecraft independently. The mapping is operated every 30 minutes using the in-situ averaged plasma properties (velocity, density, magnetic field).

\subsection{Uncertainty on the stream association from backmapping} \label{appendix:subsec_uncertainty_stream_assos}

Considering the difficulty to estimate a global backmapping uncertainty over the two steps (streamline tracing for $r>\rss$ and PFSS reconstruction for $r<\rss$), we estimate \rev{separately uncertainties} on the source alignment association \rev{and on the PFSS extrapolation.} 

\paragraph{Streamline uncertainty}
Uncertainties are estimated by the streamline deviation induced by a potential change of velocity profile due to a strong super-expansion in the corona \citep[$r<\rss$,][]{dakeyo2024b}. Such uncertainties includes exclusively a variation of the estimated solar wind velocity profile (so it does not include the PFSS uncertainty estimation). The reference isopoly profile are the "f-subsonic" type of solutions from \cite{dakeyo2022}, meaning the wind is subsonic in the super-expansion region, i.e. $\rc > \rss$ where $\rc$ is the distance at which the wind speed is crossing the sound speed. 
We suppose that the possible velocity variation is well represented by comparing the above "f-subsonic" wind to an "f-supersonic" type of solution from \cite{dakeyo2024b}. This "f-supersonic" solution has a supersonic wind flow in the super-expansion region ($\rc < \rss$).  This induces a strong difference in bulk speed radial evolution, especially above $\rss$.  The estimated deviation between f-subsonic and f-supersonic solutions is of the order of $4^\circ$ for wind speed of $350$~km/s, and there is no significant deviation for wind faster than $623$~km/s.  We summarize the results with a simple linear law~:
\begin{align}
    \dtp^{map} = \max(4 \times (623-v_{obs}) /(623-350),0) \,.
    \label{eq:uncert_dtp_mapping_dang}
\end{align}
\noindent 
The angular uncertainty $\dtp^{map}$ lies between $0^\circ$ and $4^\circ$. This defines a circular zone of uncertainty around the computed fieldline footpoints.

\paragraph{PFSS uncertainty}
Next, the uncertainty estimation on the PFSS reconstruction is very variable due to the fact that it depends on the time of observation (period of solar cycle), the temporal evolution of the source magnetic structure, the magnetic connectivity properties (in particular how much field lines diverge), the latitude of the estimated source region, the resolution of the magnetogram, the choice of the source surface height, and the choice of the reconstruction spatial resolution (order of spherical harmonics decomposition).

\rev{Therefore, an overall uncertainty relating to the PFSS cannot be quantitatively estimated as precisely as for the spiral determination uncertainty. When focusing on relatively short timescale intervals, different causes of PFSS uncertainties can be quantitatively estimated. Due to the large variety in mapped sources, estimating the uncertainty on PFSS mappings on timescales of years is difficult. We encapsulate a global uncertainty value, as estimating each individual cause of uncertainty is too complex in the present study.}

We choose to set a \rev{global} constant uncertainty on PFSS\rev{, relying on a case of a source size error}. Accounting for a relatively small to medium source size (e.g. active region angular extension), we set the PFSS uncertainty to $5^\circ$.
This add up to $\dtp^{map}$ computed above.

One might also wonder how the degree of spherical decomposition, l, affects footpoint determination. 
A study by \cite{poduval2004}, which focuses on the degree of decomposition and its impact on footpoint estimation, shows that, for $l > 22$, the estimated longitude and latitude do not vary significantly at higher orders. Regarding the value of $l = 20$ used in this study, compared to the highest order computed in \cite{poduval2004} ($l = 30$), the variation in degrees is 1–3 degrees. This is smaller than the values estimated from source size uncertainties. Thus, we have decided to consider the largest source of uncertainty, which is the source size.

\paragraph{Heliospheric current sheet influence}
\rev{Plasma parcels originating from different magnetic structures on either side of the heliospheric current sheet (HCS) are expected to embed nearby streamlines at the source surface height. This could be problematic in the streamline association process if plasma parcels originating structures with quite different source properties are associated together. 
Nevertheless, due to the large number of in-situ criteria, it is highly improbable that the source alignment method misassociates streamlines evolving nearby the HCS. Indeed, such plasma parcels are very likely to differ in intrinsic in-situ bulk properties, and their associated radial evolution. This makes the streamline association process to filter such cases.  }

\section{Source alignment and plasma identification criteria}
\label{subsec:appendix_source_align_identif_crit}
The source alignment conjunction are based on the idea to follow 
the radial evolution of similar parcels of plasma, with close intrinsic characteristics, and originating from the same solar source region. 
We recall that the source location is computed independently for PSP and SO.
The quantities referred at Parker Solar Probe and Solar Orbiter are denoted by $\mathrm{PSP}$ and $\mathrm{SO}$, respectively. Our set of criteria validating a source alignment are~: 
\begin{itemize}
    \item[-] Same magnetic polarity for the radial magnetic field $B_r$~: $\text{sign}(\Brpsp)=\text{sign}(\Brsolo)$;
    \item[-] Similar expansion factor $\fss$~: $3/4< \fpsp / \fsolo<4/3$, where $\fss$ is defined as the super-expansion of the magnetic flux tube between the photosphere and the source surface at $\rss = 2.5\, r_\odot$, deduced from the PFSS modeling by $\fss = B_r (r_\odot) \, r_\odot^2 / (B_r(\rss) \, \rss^2)$;
    \item[-] Similar photospheric radial magnetic field $B_r(r_0)$~: $1/2 < \Bropsp/\Brosolo<2$;  
    \item[-] Maximal photospheric footpoint deviation of\\ 
    $\dtp \leq 5^\circ + \dtp^{map},$ with $\dtp^{map}$ defined by Equation \eqref{eq:uncert_dtp_mapping_dang} and with $\dtp = \sqrt{\delta_\theta^2 + \cos((\thetapsp+\thetasolo)/2) \: \times \delta_\phi^2}$ where $\delta_\theta$ and $\delta_\phi$ are the difference of footpoints location in Carrington latitude and longitude, respectively. The cosine factor is there to correct the latitudinal effect that induce smaller actual displacement for the same longitude variation.
    The criteria on $\dtp$ traduces the uncertainties of the PFSS mapping process with an error on the footpoints determination of $5^\circ$, and a potential angular deviation of streamline ($\dtp^{map}$). \\
\end{itemize}

Within the potential matches, only the observations that present the closest time departure are associated, to ensure as good as possible the coherence of the temporal stream evolution. 
The parcel identification criteria are~: 
\begin{itemize}
    \item[-] Nearly spherical expansion~: $n \propto r^{-\beta}$ and $B_r \propto r^{-\delta}$ with $(\beta, \delta) \in [1.5, 2.5]$; 
    \item[-] Solar wind like temperature decrease~: $\Tp \propto r^{-\alpha}$ with $\alpha \in [0, 2]$; 
    \item[-] Similar estimated time departure from the Sun $|\tpsp - \tsolo | < 2 \, \text{days}$; the difference between the stream observation time and its estimated time travel from the solar corona \rev{is} based on the Equation~(3) of \cite{dakeyo2024a};
    \item[-] Similar mass flux  $\Mf = n \: u \: r^2 $ at the two probes with $1/2<\Mfpsp \: / \: \Mfsolo<2$; 
    \item[-] A limited change from slow to fast wind population based on the five winds speed classification, i.e. evolution of maximal 2 wind population up or down between Parker Solar Probe and Solar Orbiter wind populations;   
    \item[-] The average bulk velocity changes $\Dv$ per radial decade (Equation~\eqref{eq:Delta_v_express}) deviating too much from the referent average $\Dv$ are discarded. More precisely the average $\Dv$ is derived from the unclassified join dataset of PSP and Solar Orbiter (computed from a logarithmic fit of the form $v = a \times \log(r) + b$ where $a$ and $b$ two constants). 
    The previously mentioned referent average speed increase per decade $\Dv_{ref} = 128$~km/s, is used as a limit for stream association such as $-\Dv_{ref} < \Dv < +2 \: \Dv_{ref}$. 
    This criteria is related to the above criteria on limiting the wind population change between PSP and SO. 
\end{itemize}

We can see that some of the criteria have relatively broad constraints, particularly with regard to the parcel identification criteria for temperature, density and magnetic field radial evolution. These broad criteria aim to exclude non-solar wind-like evolution while also avoiding evolution to be too constrained. While previous statistical studies can provide expected trends \citep{maksimovic2020, halekas2022, dakeyo2022}, the radial evolution of single solar wind streams is still poorly understood. Therefore, we do not want to introduce bias into the stream association by setting strict criteria according to other solar wind studies results.
Moreover, while each broad criterion alone is insufficient to conduct a reliable association, accounting for a large set of these criteria excludes extreme cases step by step. Many streams observed between PSP and SO can satisfy a single criterion alone (especially with broad limits), but very few satisfy all of them.

Regarding previous study in the literature \citep{Telloni2021, Ervin2024a, Ervin2024b, Berriot2024, Rivera2024, Rivera2025}, while the backmapping method used for source alignment is in essence similar to that of individual Parker spiral alignment, the statistical stream-by-stream association process and the analysis of these results, are based on a different methodology to that described in the literature. Moreover, in this study PSP and SO respective observations are spread over relatively large radial distances, making individual stream's radial coverage to differ from one to another.

\end{document}